%% file: main.tex
\documentclass{llncs}
\usepackage{wrapfig}
\usepackage[ruled,linesnumbered,noend]{algorithm2e}
\usepackage{citesort}
\usepackage{multicol}
\usepackage{amssymb}
\usepackage{amsmath,bbm}
\usepackage{latexsym,epic,eepic}
\usepackage{complexity}
\usepackage{multirow}
\usepackage{algorithmic}
\usepackage{url}
\usepackage{btran}
\usepackage{times}
\usepackage{tikz}
\usetikzlibrary{shapes}
\usetikzlibrary{automata,positioning}

\include{comm-anthony}
\include{comm}

\SetCommentSty{mycommfont}
\newlength{\commentWidth}
\newcommand{\atcp}[1]{\tcp*[r]{\makebox[\commentWidth]{#1\hfill}}}

\newif\ifdraft\draftfalse
\ifdraft
\newcommand\anthony[1]{{\color{blue}
[#1 - \textbf{Anthony}]}}
\newcommand\philipp[1]{{\color{magenta}
[#1 - \textbf{Philipp}]}}

\newcommand\anthonychanged[1]{{\color{red}{#1}}}

\newcommand\todo[1]{}
\else
\newcommand\todo[1]{}
\newcommand\anthony[1]{}
\newcommand\philipp[1]{}
\newcommand\khanh[1]{}
\newcommand\sunjun[1]{}

\newcommand\anthonychanged[1]{#1}

\fi

\newcommand\shortlong[2]{#2}

\usepackage{color}
\shortlong{}{\pagestyle{plain}}

\title{Liveness of Randomised Parameterised Systems\\
    under Arbitrary Schedulers\shortlong{}{
(Technical Report)
}}
\author{ 
    Anthony W. Lin\inst{1} \and
    Philipp R\"ummer\inst{2} 
}
\institute{
    Yale-NUS College, Singapore \and
    Uppsala University, Sweden 
}
\date{}

\begin{document}

\maketitle

\vspace*{-4ex}
\begin{abstract}
    \input{abstract}

\end{abstract}

\input{introduction}
\input{prelim}

\input{framework}
\input{autoReachability}
\input{experiments}
\input{conc}

\paragraph{Acknowledgment.} We thank anonymous referees, Parosh Abdulla,
Bengt Jonsson, Ondrej Lengal, Rupak Majumdar, and Ahmed Rezine for their helpful
feedback. We 
thank Truong Khanh Nguyen for contributing with the development of the tool 
\texttt{parasymmetry} 
\cite{DBLP:conf/vmcai/LinNR016}, on top of which our current tool (SLRP) 
builds.

\anthony{Need to shorten the references. Looks ugly as they are now.}

\bibliographystyle{abbrv}
\bibliography{references}

\shortlong{}{
\appendix

\clearpage
\begin{center}
    \LARGE \bfseries APPENDIX
\end{center}

\input{app}

}

\end{document}

%% file: comm-anthony.tex
\newcommand{\Props}{\ensuremath{\text{AP}}}
\newcommand{\OMIT}[1]{}
\newcommand{\ACT}{\ensuremath{\text{\sf ACT}}}

\newcommand{\defn}[1]{\textit{#1}} 
\newcommand{\mcl}[1]{\ensuremath{\mathcal{#1}}}
\newcommand{\sem}[1]{[\![#1 ]\!]}

\newcommand{\struct}{\mathfrak{S}}
\newcommand{\Tree}{\ensuremath{\text{\scshape Tree}}}

\newcommand{\transys}{\ensuremath{\langle S; \to\rangle}}
\newcommand{\transysMDP}{\ensuremath{\langle S; \to_1, \to_2 \rangle}}
\newcommand{\transysR}{\ensuremath{\langle S; R\rangle}}
\newcommand{\transysG}{\ensuremath{\langle S; \{\to_a\}_{a \in \ACT}\rangle}}
\newcommand{\Kripke}{\ensuremath{\langle S; \{\to_a\}_{a \in \ACT}, \{U_b\}_{b
\in \Props}\rangle}}

\newcommand{\Rec}{\ensuremath{\text{\scshape Rec}}}

\newcommand{\empseq}{\ensuremath{\epsilon}}

\newcommand{\ialphabet}{\ensuremath{\Sigma}}

\newcommand{\dalphabet}{\ensuremath{\Omega}}

\newcommand{\N}{\ensuremath{\mathbb{N}}}

\newcommand{\Z}{\ensuremath{\mathbb{Z}}}

\newtheorem{convention}{Convention} 

\newcommand{\problemx}[3]{
\par\noindent\underline{\sc#1}\par\nobreak\vskip.2\baselineskip
\begingroup\clubpenalty10000\widowpenalty10000
\setbox0\hbox{\bf Instance: }\setbox1\hbox{\bf Question: }
\dimen0=\wd0\ifnum\wd1>\dimen0\dimen0=\wd1\fi
\vskip-\parskip\noindent
\hbox to\dimen0{\box0\hfil}\hangindent\dimen0\hangafter1\ignorespaces#2\par
\vskip-\parskip\noindent
\hbox to\dimen0{\box1\hfil}\hangindent\dimen0\hangafter1\ignorespaces#3\par
\endgroup}

\newcommand{\Paths}{\ensuremath{\text{\scshape Paths}}}

\newcommand{\Aut}{\ensuremath{\mathcal{A}}} 
\newcommand{\AutABA}{\ensuremath{\Aut_A}} 
\newcommand{\Tra}{\ensuremath{\mathcal{T}}} 
\newcommand{\TraPrec}{\ensuremath{\Tra_\prec}} 
\newcommand{\transrel}{\ensuremath{\Delta}} 
\newcommand{\tran}[1]{\ensuremath{\stackrel{#1}{\longrightarrow}}}
\newcommand{\controls}{\ensuremath{Q}} 
\newcommand{\finals}{\ensuremath{F}} 
\newcommand{\Lang}{\ensuremath{\mathcal{L}}} 
\newcommand{\AutRun}{\ensuremath{\rho}} 

\newcommand{\arity}{\ensuremath{\text{\texttt{ar}}}}

\newcommand{\rarw}{\rightarrow}

\newcommand{\ModelRun}{\ensuremath{\pi}}

\newcommand{\Size}[1]{\ensuremath{\|#1\|}}

\newcommand{\DEC}{\ensuremath{\text{\texttt{DEC}}}}
\newcommand{\INC}{\ensuremath{\text{\texttt{INC}}}}

\newcommand{\Prob}{\ensuremath{\text{Prob}}}

\newcommand{\LeftW}{\ensuremath{\stackrel{\longleftarrow}{W}}}
\newcommand{\RightW}{\ensuremath{\stackrel{\longrightarrow}{W}}}
\newcommand{\LeftS}{\ensuremath{\stackrel{\longleftarrow}{S}}}
\newcommand{\RightS}{\ensuremath{\stackrel{\longrightarrow}{S}}}
\newcommand{\LeftD}{\ensuremath{\stackrel{\longleftarrow}{D}}}
\newcommand{\RightD}{\ensuremath{\stackrel{\longrightarrow}{D}}}
\newcommand{\LeftHold}{\ensuremath{\stackrel{\longleftarrow}{hold}}}
\newcommand{\RightHold}{\ensuremath{\stackrel{\longrightarrow}{hold}}}

%% file: comm.tex
\newcommand{\biword}[2]{(#1,#2)}

%% file: abstract.tex
We consider the problem of verifying liveness for systems with a finite,
but unbounded, number of processes, commonly known as \emph{parameterised
systems}. Typical examples of such systems include distributed protocols (e.g. 
for the dining philosopher problem).
Unlike the case of verifying safety,
proving liveness is still considered extremely challenging,
especially in the presence of randomness in the system.
In this paper we consider liveness under arbitrary (including unfair)
schedulers, which is often considered a desirable property in the literature
of self-stabilising systems.
We introduce an automatic method of proving liveness 
for randomised parameterised systems under arbitrary schedulers. 
Viewing liveness as a two-player reachability game (between Scheduler and 
Process), our method is a CEGAR approach 
that synthesises a progress relation for Process that can be symbolically 
represented as a finite-state automaton. The method is incremental and
exploits both Angluin-style L*-learning and SAT-solvers.
Our experiments show that our algorithm is able to prove liveness 
automatically for well-known randomised distributed protocols, including 
Lehmann-Rabin Randomised Dining Philosopher Protocol and 
randomised self-stabilising protocols (such as the Israeli-Jalfon Protocol).
To the best of our knowledge, this is the first fully-automatic method that can 
prove liveness for randomised protocols.

%% file: introduction.tex
\section{Introduction}
\label{sec:intro}

Verification of parameterised systems is one of the most extensively
studied problems in computer-aided verification. Parameterised systems
are infinite families of finite-state systems that are described in some
finite behavioral description language. Distributed protocols (e.g. for the 
dining philosopher problem) are typical examples of parameterised systems since
they can represent any finite (but unbounded) number of processes. Verifying 
a parameterised system, then, amounts to verifying \emph{every instance} of
the infinite family. In the case of a dining philosopher protocol, this 
amounts to verifying the protocol with any number of philosophers. Although the
problem was long known to be undecidable \cite{AK86}, a lot of progress
has been made to tackle the problem resulting in such techniques as network
invariants (including cutoff techniques), symbolic model checking
(including regular model checking), and finite-range abstractions, to name
a few.
The reader is referred to the following excellent surveys
\cite{vojnar-habilitation,rmc-survey,Zuck-survey,sasha-book,Parosh12} 
covering these different approaches to solving the problem.

Nowadays there are
highly effective automatic methods that can successfully verify 
\emph{safety} for many
parameterised systems derived from real-world concurrent/distributed
algorithms (e.g. see
\cite{Lever,AHH13,BFLP08,RMC-without,RMC,WB98,BHRV12,BHV04,Neider13,KKW10,trex,ADR07,Parosh12,vojnar-habilitation,BLW03,TORMC,vojnar-habilitation,rmc-survey,forest-automata,Lin12-fsttcs,AAC13,HL12,EGP14}).
In contrast, there has been much less progress in automatic
techniques for proving \emph{liveness} for 
parameterised systems. In fact, this difficulty has also been widely observed 
(e.g. see \cite{PS00,vojnar-habilitation,JS07,AJRS06}).
Proving liveness amounts to proving that,
under a class of adversarial schedulers (a.k.a. \emph{adversaries} or just 
\emph{schedulers}), something ``good'' will eventually happen.
The problem is known to be reducible to 
finding an infinite path satisfying a B\"{u}chi condition (e.g. see 
\cite{vojnar-habilitation,rmc-survey,PS00,TL08,rmc-thesis,handling-liveness,Lever-omega,TL10}).
The latter problem (a.k.a. repeated reachability) in general requires
reasoning about the transitive closure relations,
which are generally observed to be rather difficult to compute automatically.

\emph{Randomised parameterised systems} are infinite families of finite-state
systems that allow both nondeterministic and probabilistic transitions
(a.k.a. Markov Decision Processes \cite{marta-survey}). 
This paper concerns the problem of verifying liveness for
randomised parameterised systems, with an eye towards a
fully-automatic verification algorithm for well-known \emph{randomised 
distributed protocols} that commonly feature in finite-state probabilistic
model checkers (e.g. PRISM \cite{KNP11}), but have so far resisted
fully-automatic parameterised verification.
Such protocols include Lehmann-Rabin's Randomised Dining Philosopher Protocol 
\cite{LR81} and randomised self-stabilising protocols (e.g. Israeli-Jalfon's 
Protocol \cite{IJ90} and Herman's Protocol \cite{Her90}), to name a few. 
Randomised protocols generalise deterministic protocols by allowing each process
to make probabilistic transitions, i.e., not just a transition with
probability 1.
Randomisation is well-known to be useful in the design of distributed protocols,
e.g., to break symmetry and simplifies distributed algorithms (e.g. see
\cite{Lynch-book,Fokkink-book}). 
Despite the benefits of randomisation in protocol design, the use of
randomisation makes proving liveness substantially more challenging (e.g. see 
\cite{LSS94,PZ86,Lynch-book}).
Proving liveness for probabilistic distributed protocols amounts to
proving that, under a class of adversaries, something ``good'' will eventually
happen \emph{with probability 1} (e.g. see
\cite{Lynch-book,CY95,Var85,LS07,marta-survey,APZ03,EGK12}). 
\OMIT{
[For this
reason, the property is sometimes referred to as 
}
\OMIT{
For example, Lehmann-Rabin protocol satisfies \emph{freedom
from (individual) starvation} under fair schedulers \cite{Lynch-book} 
(i.e. every philosopher will eventually eat). [In fact, some liveness condition (e.g. \emph{some} 
philosopher will eventually eat) is still satisfied under all
(possibly unfair) schedulers, provided that a philosopher is not allowed to
make an idle transition when chosen by a scheduler \cite{DFP04}.]
}
Unlike the case of deterministic protocols, proving liveness for
probabilistic protocols requires reasoning about \emph{games} between 
an adversary and a stochastic process player (a.k.a. $1\frac{1}{2}$-player 
game), which makes the problem computationally more difficult even 
in the finite-state case (e.g. see \cite{LS07}). To the best of our knowledge,
there is presently no fully-automatic technique which can prove liveness
for such randomised distributed protocols as Lehmann-Rabin's Randomised
Distributed Protocols \cite{LR81}, and self-stabilising randomised protocols 
including Israeli-Jalfon's Protocol \cite{IJ90} and Herman's Protocol 
\cite{Her90}.

\medskip
\noindent
\textbf{Contribution:}
The main contribution of the paper is a fully-automatic 
method for proving liveness over randomised parameterised systems 
over various network topologies (e.g. lines, rings, stars, and cliques) 
under arbitrary (including unfair) schedulers. Liveness under arbitrary 
schedulers is a desirable property in the literature of self-stabilising
algorithms
since an unfair scheduler (a.k.a. daemon) enables a \emph{worst-case} analysis
of an algorithm and covers the situation when some process is ``frozen'' 
due to conditions that are external to the process (e.g. see 
\cite{GS13,BGJ07,taxonomy-daemons,KY97}). There are numerous examples of
self-stabilising protocols that satisfy liveness even under unfair schedulers
(e.g. see \cite{Dijkstra74,Fokkink-book,IJ90,BGJ07,KY97}). Similar examples are
also available in the literature of mutual exclusion protocols (e.g. 
\cite{DFP04,Szymanski}), and consensus/broadcast protocols
(e.g. \cite{Fokkink-book,BT85}).
Our algorithm can successfully verify liveness
under arbitrary schedulers for 
a fragment of FireWire's symmetry breaking protocol \cite{EGK12,MMH03},
Israeli-Jalfon's Protocol \cite{IJ90}, 
Herman's Protocol \cite{Her90} considered over a linear array, and
Lehmann-Rabin Dining Philosopher Protocol \cite{DFP04,LR81}.

It is well-known that for proving liveness for a finite-state Markov Decision
Process (MDP) only the topology of the system matters, not
the actual probability values (e.g. see \cite{CY95,Var85,Alfaro99,HSP83}).
Hence, the same is true for randomised parameterised systems since each
instance is a finite MDP. In this paper, we follow this approach and view
the problem of proving liveness under arbitrary schedulers as a 2-player 
reachability game between Scheduler (Player 1) and Process (Player 2) over 
non-stochastic parameterised systems, obtained by simply ignoring the actual
probability values of transitions with non-zero probabilities (transitions with 
zero probability are removed). This simple reduction allows us to adopt any
symbolic representation of non-stochastic parameterised systems.
In this paper, we represent parameterised systems as finite-state
letter-to-letter transducers, as is standard in \emph{regular model checking} 
\cite{vojnar-habilitation,rmc-survey,rmc-thesis,RMC,Parosh12}. In this 
framework, 
configurations of parameterised systems are represented as words over a 
finite alphabet $\ialphabet$ (usually encoding a finite set of control
states for each local process). Many distributed protocols that arise in 
practice can be naturally modelled as 
transducers.

To automatically verify liveness of parameterised systems in this
representation, we develop a counterexample-guided method for
synthesising Player~2 strategies. The core step of the approach is the
computation of \emph{well-founded relations} guiding Player~2 towards
winning configurations (and the system towards ``good'' states). In
the spirit of regular model checking, such well-founded relations are
represented as letter-to-letter transducers; however, unlike most
regular model checking algorithms, we use learning and SAT-based
methods to compute the relations, in line with some of the recent
research on the application of learning for program analysis
(e.g. \cite{Neider10,Neider13,DBLP:conf/cav/0001LMN13,NT16}). This
gives rise to a counterexample-guided algorithm for computing winning
strategies for Player~2.  We then introduce a number of refinements of
the base method, which turn out to be essential for analysing
challenging systems like the Lehmann-Rabin protocol: strategies for
Player~2 can be constructed \emph{incrementally,} reducing the size of
automata that have to be considered in each inference step; symmetries
of games (e.g., rotation symmetry in case of protocols with ring
topology) can be exploited for acceleration; and inductive
over-approximations of the set of reachable configurations can be
pre-computed with the help of learning. To the best of our knowledge,
the last refinement also represents the first successful application
of Angluin's L*-algorithm~\cite{Angluin:1987:LRS:36888.36889} for
learning DFAs representing inductive invariants in the regular model
checking context.

\OMIT{
Our method for proving liveness for probabilistic paramterised systems
is based on synthesising \emph{advice bits} (in the form of 
finite-state automata or transducers) that witness a win for Player 2 under 
optimal plays. Intuitively, advice bits are an \emph{overapproximation} of the 
winning strategy for Player 2. Verifying advice bits actually witness a win
for Player 2 under optimal plays can be done fully-automatically using an
automata-based technique, e.g., within the framework of \emph{automatic 
transition systems} \cite{BG04,Blum99}.
However, a naive enumeration of advice bits is impractical. For this reason,
we provide a CEGAR method for synthesising advice bits. The CEGAR method
will sequentially go through $k=1,2,\ldots$ and attempts to synthesise advice 
bits whose automata/transducers contain only $k$ states. To make the
approach viable, we encode candidate advice bits as satisfying assignments to 
constraints $\varphi$ in propositonal logic, which are generated by a highly 
optimised SAT-solver. 
Whether candidate advice bits actually witness a win for Player 2 under
optimal plays are then checked 
using an automata-based method. In the case when they do not witness a win
for Player 2, a counterexample (in terms of tuple of words) is generated, which 
is then incorporated into the constraint $\varphi$ and used
to refine the subsequent guess by a SAT-solver.
}

We have implemented our method as a proof of concept. Besides the 
four aforementioned probabilistic protocols that we have successfully verified 
against liveness (under all schedulers), we also show that our tool is 
competitive with existing tools (e.g. \cite{AJRS06,rmc-thesis}) for proving 
liveness for
deterministic parameterised systems (Szymanski's mutual exclusion protocol
\cite{Szymanski}, Left-Right Dining Philosopher Protocol \cite{Lynch-book},
Lamport's Bakery Algorithm \cite{Ben-Ari,Fokkink-book}, and Resource-Allocator
Protocol \cite{donaldson-thesis}). Finally, we
report that our tool can also automatically solve classic
examples from 
combinatorial game theory on infinite graphs
(take-away game and 
Nim~\cite{Ferguson}). To the best
of our knowledge, our tool is the first verification tool that can automatically
solve these games.

\medskip
\noindent
\textbf{Related Work:}
There are currently only a handful of fully-automatic techniques 
for proving liveness for randomised parameterised systems. We mention 
the works 
\cite{EGK12,Sriram13,Mon01} on proving almost-sure termination of sequential 
probabilistic programs. Strictly speaking, these works are not directly 
comparable to our
work since their tools/techniques handle only programs with variables
over integer/real domains, and cannot naturally model the protocol examples over
line/ring topology that we consider in this paper. 
Based on the work of Arons \emph{et
al.} \cite{APZ03}, the approach of Esparza \emph{et al.} \cite{EGK12} aims to 
guess a terminating pattern by constructing a nondeterministic program from
a given probabilistic program and a terminating pattern candidate. This allows
them to exploit model checkers and termination provers for nondeterministic 
programs. The approach is sound and
complete for ``weakly-finite'' programs, 
which include parameterised programs, i.e., programs with parameters that can 
be initialised to arbitrary large values, but are finite-state for every 
valuation of the parameters. 
The approach of \cite{Sriram13} is a constraint-based  method
to synthesise ranking functions for probabilistic programs based on 
martingales and may be able to prove almost sure termination for probabilistic 
programs that are not weakly finite. Monniaux \cite{Mon01} proposed
a method for proving almost sure termination for probabilistic programs
using abstract interpretation, though without tool support.


As previously mentioned, there is a lot of work on liveness for 
non-probabilistic parameterised systems (e.g. see
\cite{PXZ02,rmc-thesis,AJRS06,PS00,TL08,Lever-omega,handling-liveness,liveness-invisible,TL10}).
We assess our technique in this context by
using several typical benchmarking examples that satisfy liveness (more
precisely, deadlock-freedom) under arbitrary schedulers including Szymanski's 
Protocol, Bakery Protocol, and Deterministic Dining Philosopher with 
Left-Right Strategy. 

Two-player reachability games on automatic graphs (i.e. regular model checking
with non-length preserving transducers) have been considered by Neider 
\cite{Neider10},
who proposed an L*-based learning algorithm for constructing the set of 
winning regions enriched with ``distance'' information, which is a number 
that can be represented in binary or unary. [Embedding distance information in 
a reachability set was first done in regular model checking by Vardhan 
\emph{et al.} \cite{Lever}] Augmenting winning regions or reachability sets with distance 
information, however, often makes regular sets no longer regular 
\cite{Neider13}.
In this paper, we do not consider non-length preserving transducers and our
algorithm is based on constructing progress relations for Player 2. 
In particular, part of our algorithm employs an L*-based algorithm
for synthesising an inductive invariant which, however, differs from
\cite{Neider10,Lever} since membership tests (i.e. reachability of a single 
configuration) are decidable.
Recently
Neider and Topcu \cite{NT16} proposed a learning algorithm for solving safety 
games over rational graphs (an extension of automatic graphs), which are 
\emph{dual} to reachability games.

%% file: prelim.tex
\section{Preliminaries}
\label{sec:prelim}

\noindent
\textbf{General notations}:
For any two given real numbers $i \leq j$, we use a standard notation
(with an extra subscript) to denote real intervals, e.g., $[i,j]_{\mathbb{R}} = 
\{ k \in \mathbb{R} : i \leq k \leq j \}$ and $(i,j] \{ k \in \mathbb{R} : i < 
k \leq j \}$. We will denote intervals over integers by removing the
subscript, e.g., $[i,j] := [i,j]_{\mathbb{R}} \cap \Z$.
Given a set $S$, we use $S^*$ to denote the set of all finite
sequences of elements from $S$. The set $S^*$ always includes the empty
sequence which we denote by $\empseq$. 
Given two sets of words
$S_1, S_2$, we use $S_1\cdot S_2$ to denote the set $\{ v\cdot w: v\in S_1,
w\in S_2\}$ of words formed by concatenating words from $S_1$ with words from
$S_2$. Given two relations $R_1,R_2 \subseteq S \times S$, we define
their composition as $R_1 \circ R_2 = \{ (s_1,s_3) : (\exists s_2)((s_1,s_2)
\in R_1 \wedge (s_2,s_3) \in R_2)\}$. 
\smallskip

\noindent
\textbf{Transition systems}: Let $\ACT$ be a finite set of
\defn{action symbols}. 
A \defn{transition system} over $\ACT$ is
a tuple $\struct = \Kripke$,
where $S$ is a set of \defn{configurations}, $\to_a\ \subseteq S \times S$ 
is a binary relation over $S$, and $U_b \subseteq S$ is a unary relation on
$S$. 
In the sequel, we will often consider transition systems where $\Props = 
\emptyset$ and $|\ACT| = 1$, in which case $\Kripke$ will be denoted as
$\transys$. If $|\ACT| > 1$, we use $\to$ to denote the relation 
$\left(\bigcup_{a \in \ACT} \to_a\right)$. 
The notation $\to^+$ (resp. $\to^*$) is used to denote the transitive (resp.
transitive-reflexive) closure of $\to$. 
We say that a sequence $s_1 \to \cdots \to s_n$ is a \defn{path} (or
\defn{run}) in $\struct$ (or in $\to$). Given two paths $\ModelRun_1: s_1
\to^* s_2$ and $\ModelRun_2: s_2 \to^* s_3$ in $\to$, we may concatenate them
to obtain $\ModelRun_1 \odot \ModelRun_2$ (by gluing together $s_2$).
We call $\ModelRun_1$ a \defn{prefix} of $\ModelRun_1 \odot \ModelRun_2$.
For each $S' \subseteq S$, we use the notations $pre_{\to}(S')$ and 
$post_{\to}(S')$ to denote the pre/post image of $S'$ under $\to$.
That is, $pre_{\to}(S') := \{ p \in S : \exists q\in S'( p \to q ) \}$ and
$post_{\to}(S') := \{ q \in S : \exists p\in S'( p \to q ) \}$.

\OMIT{
When dealing with probabilistic systems, we will find the following notations
handy. For two sets $S, S' \subseteq S$ of configurations in $\struct$, denote
by $\Paths_\struct(S,S')$ the set of all paths from (some state in) $S$ to 
(some state in) $S'$. We will omit mention of $\struct$ if $\struct$ is
clear from the context.
\anthony{Need to check if we can remove this notation}
}
\OMIT{
Given a relation $\to \subseteq S \times S$ and subsets 
$S_1,\ldots,S_n \subseteq S$, denote by 
$\Rec_{\to}(\{S_i\}_{i=1}^n)$ to be the set of elements $s_0 \in S$ for which
there exists an infinite path $s_0 \to s_1 \to \cdots$ visiting
each $S_i$ infinitely often, i.e., such that, for each
$i\in[1,n]$, there are infinitely many $j \in \N$ with $s_j \in S_i$.
}
\smallskip


%

\noindent
\textbf{Words and automata}:
We assume basic familiarity with word automata.
Fix a finite alphabet $\Sigma$. For each finite word $w = w_1\ldots w_n \in 
\Sigma^*$, we
write $w[i,j]$, where $1 \leq i \leq j \leq n$, to denote the segment
$w_i\ldots w_j$. Given an automaton $\mcl{A} = (\Sigma,Q,\delta,q_0,F)$,
a run of $\mcl{A}$ on $w$ is a function $\rho: \{0,\ldots,n\}
\rarw Q$ with $\rho(0) = q_0$ that obeys the transition relation $\delta$. 
We may also denote the run $\rho$ by the word $\rho(0)\cdots \rho(n)$ over
the alphabet $Q$. 
The run $\rho$ is said to be \defn{accepting} if $\rho(n) \in F$, in which
case we say that the word $w$ is \defn{accepted} by $\mcl{A}$. The language
$L(\mcl{A})$ of $\mcl{A}$ is the set of words in $\Sigma^*$ accepted by
$\mcl{A}$.

\smallskip
\noindent
\textbf{Reachability games}:
We recall some basic concepts on 2-player reachability games 
(e.g. see \cite[Chapter 2]{ALG-book} on games with 1-accepting 
conditions).
An \defn{arena} is a transition system $\struct = \transysMDP$, where $S$
(i.e. the set of ``game configurations'')
is partitioned into two disjoint sets $V_1$ and $V_2$ such that $pre_{\to_i}(S)
\subseteq V_i$ for each $i=1,2$. The transition relation $\to_i$ 
denotes the actions of Player $i$.
Similarly, for each $i=1,2$, the configurations $V_i$ are controlled by 
Player $i$. In the sequel, Player 1 will
also be called ``Scheduler'', and Player 2 ``Process''. Given a set
$I_0 \subseteq S$ of initial states and a set $F \subseteq S$ of 
final (a.k.a. target) states, the goal of Player 2 is to reach $F$
from $I_0$,
while the goal of Player 1 is to avoid it. 
More formally, a \defn{strategy} for Player $i$ is a 
partial function $f: S^*V_i \to S$ such that, for each $v \in S^*$ and
$p \in V_i$, if $vp$ is a path in $\struct$ and that $p$ is not a dead end 
(i.e. $p \to_i q$ for some $q$), then
$f(vp)$ is defined in such a way that $p \to_i f(vp)$. Given a strategy $f_i$
for Player $i=1,2$ and an initial state $s_0 \in S$, we can define a unique
(finite or infinite) path in $\struct$ 
$\pi: s_0 \to_{j_1} s_1 \to_{j_2} \cdots$
such that $s_{j_{k+1}} = f_i(s_0s_1\ldots s_{j_k})$ where $i \in \{1,2\}$
is the (unique) number such that $s_{j_k} \in V_i$. 
Player 2 \emph{wins} iff some state 
in $F$ appears in $\pi$, or if the path is finite and the last configuration
belongs to Player 1. Player 1 \emph{wins} iff Player 2 does not
win (i.e. \emph{loses}). A strategy $f$ for Player $i$ 
is \defn{winning} from $I_0$, for each strategy 
$g$ for Player $i+1 \pmod{2}$, the unique path in $\struct$ from each 
$s_0 \in I_0$
witnesses a win for Player $i$. Such games (a.k.a. \emph{reachability games})
are \emph{determined} (e.g. see \cite[Proposition 2.21]{ALG-book}), 
i.e., either Player 1 has a winning strategy or Player 2 has a winning strategy.


%


%
\OMIT{
later, to deal with
liveness over parameterised systems it suffices to deal with the following
computational problem for reachability games: given an arena $\struct = 
\transysG$, a set $I_0 \subseteq S$ of initial configurations, 
and a set $F \subseteq S$ of final configurations, decide whether
Player 2 has a winning strategy from \emph{all of} $I_0$ to reach $F$ in 
$\struct$? [In other words, whether $I_0$ is a subset of the winning 
configurations for Player 2.] 
}

\begin{convention}
    For simplicity's sake, we make the following assumptions on our reachability
    games. They suffice for the purpose of proving liveness for parameterised
    systems. The techniques can be easily adapted when these assumptions are
    lifted.
    \begin{description}
        \item[(A0)] Arenas are \defn{strictly alternating}, i.e., 
        a move made by a player does not take the game back to her 
        configuration (i.e.  $post_{\to_i}(S) \cap A_i = \emptyset$, for each 
        $i \in \{1,2\}$).
    \item[(A1)] 
        Initial and final configurations belong to
        Player 1, i.e., $I_0, F \subseteq V_1$.
    \item[(A2)] Non-final configurations are no dead ends, i.e.,
        $\forall x\in S\setminus F, \exists y: x \to_1 y \vee x \to_2 y$.
    \end{description}
    \label{con:games}
\end{convention}

%% file: framework.tex
\section{The formal framework} \label{sec:framework}

Parameterised systems are an infinite family $\mathcal{F} = 
\{\struct_i\}_{i\in \N}$ of finite-state transition systems.
Similarly, \defn{randomised parameterised systems} are an infinite 
family $\mathcal{F} = \{\struct_i\}_{i\in \N}$ of 
\emph{Markov Decision Processes} \cite{marta-survey}, which are finite-state 
transition systems $\struct = \transysMDP$ that have both ``nondeterministic'' 
transitions $\to_1$ and ``probabilistic'' transitions $\to_2$. 

We first informally illustrate the concept of randomised parameterised systems
by means of Israeli-Jalfon Randomised Self-Stabilising Protocol \cite{IJ90}
(also see \cite{Norman04}). 
\begin{wrapfigure}{r}{24mm}
    \vspace{-10mm}
\begin{center}
    \includegraphics[width=24mm]{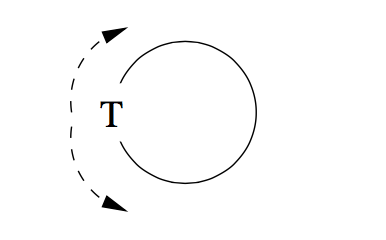}
\end{center}
    \vspace{-9mm}
\end{wrapfigure}
The protocol has a ring topology
and each process either holds a token (denoted by $\top$) or
does not hold a token (denoted by $\bot$). At any given step, the Scheduler 
chooses a process $P$ that holds a token. The process $P$ can then pass the
token to its left or right neighbour each with probability 0.5. In doing so,
two tokens that are held by a process are merged into one token (held
by the same process). It can be proven that under arbitrary schedulers, 
starting from any configuration
with \emph{at least} one token, the protocol will converge to a 
configuration with \emph{exactly} one token with probability 1. This is
an example of liveness under arbitrary schedulers.

It is well-known that the liveness problem for finite MDPs 
$\struct$ depends on the topology of the graph $\struct$, not on the
actual probability values in $\struct$ (e.g. \cite{CY95,Var85,Alfaro99,HSP83}).
In fact, this result easily transfers to randomised parameterised systems since
\emph{every} instance in the infinite family is a finite MDP. 
Following this approach, we may view the problem of proving (almost-sure) 
liveness for randomised parameterised systems under 
arbitrary schedulers as a 2-player reachability game 
between Scheduler 
(Player 1 with moves $\to_1$) and Process (Player 2 with moves $\to_2$)
over the arena $\struct = \transysMDP$ 
obtained by simply \emph{ignoring} the actual probability values of transitions 
in $\to_2$ (with non-zero probabilities). This simple reduction allows us 
to view randomised parameterised systems as an \emph{infinite family of
finite arenas} and adopt standard symbolic representations of non-stochastic 
parameterised systems  (many of which are known).
Our formal framework uses the standard symbolic representation using
letter-to-letter transducers.
To simplify our presentation, \emph{we will directly define
    liveness for randomised parameterised systems in terms of non-stochastic
two player games and relegate this standard reduction in the
\shortlong{full version}{appendix} for
interested readers.}

\subsection{Liveness as games}
Given a randomised parameterised system $\mathcal{F} = \{\struct_i\}_{i \in
\N}$, a set $I_0 \subseteq V_1$ of initial states, and a set $F \subseteq V_1$ 
of final states, we say that a randomised parameterised system \emph{satisfies 
    liveness under arbitrary schedulers with probability 1 (a.k.a. almost surely
terminates)} if from \emph{each} configuration $s_0 \in 
post_{\rightarrow^*}(I_0)$, 
Player 2 has a winning strategy reaching $F$ in $\mathcal{F}$ (viewed as an 
arena). The justification of this definition is in 
\shortlong{the full version}{Prop.~\ref{prop:removeProb}  (Appendix)}.

\OMIT{
For this reason,
we will simply state the resulting decision problem on 
}

\OMIT{
One of the corollaries of Proposition \ref{prop:removeProb} is that, as far as 
answering liveness is concerned, we may use any known symbolic representations 
of parameterised systems (many of which are known) to represent probabilistic 
parameterised systems. Our formal framework uses the standard symbolic 
representation of parameterised systems from regular model checking
\cite{rmc-survey,vojnar-habilitation,RMC,rmc-thesis}, i.e., transducers.
Many distributed protocols that arise in practice can be naturally modelled as 
transducers. 
}

\subsection{Representing infinite arenas}
Our formal framework uses the standard symbolic 
representation of parameterised systems from regular model checking
\cite{rmc-survey,vojnar-habilitation,RMC,rmc-thesis}, i.e., transducers.
Many distributed protocols that arise in practice can be naturally modelled as 
transducers. 
\defn{Transducers} are {\em letter-to-letter automata} that accept 
$k$-ary relations over words (cf.~\cite{BG04}).
In this paper, we are only interested in binary \emph{length-preserving 
relations} \cite{rmc-survey},
i.e., a relation $R \subseteq \Sigma^* \times \Sigma^*$ such that each
$(v,w) \in R$ implies that $|v| = |w|$. For this reason, we will only define
length-preserving transducers and only for the binary case.
Given two words
$w = w_1\ldots w_n$ and $w' = w_1'\ldots w_n'$ over the alphabet $\Sigma$, we
define a word $w \otimes w'$ 
over the alphabet $\Sigma \times \Sigma$ as $\biword{w_1}{w_1'}\cdots 
\biword{w_n}{w_n'}$.
\OMIT{
$$ w \otimes w' \ = \ \biword{a_1}{b_1}\ldots\biword{a_k}{b_k}, \
  \text{where}\
a_i \ = \ \begin{cases} w_i & i \leq n \\
                         \bot &  i > n,
            \end{cases}
\ \text{and} \ 
b_i \ = \ \begin{cases} w_i' & i \leq m \\
                         \bot & i > m.
            \end{cases}
$$
}
\OMIT{
In other words, the shorter word is padded with $\bot$'s, and the
$i$th letter of $w \otimes w'$ is then the pair of the $i$th letters of
padded $w$ and $w'$. 
}
A letter-to-letter transducer is simply an automaton over $\Sigma\times\Sigma$, 
and a binary relation ${R}$ over $\Sigma^*$ is \defn{regular} if
the set $\{ w \otimes w' : (w,w') \in {R} \}$ is accepted by
a letter-to-letter automaton $\mcl{R}$. Notice that the resulting relation $R$ 
only relate words that are of the same length.
In the sequel, 
to avoid notational clutter, 
we will use $R$ to mean both a transducer and the binary relation
that it recognises.

\begin{definition}[Automatic systems]
    A system $\struct =\Kripke$ is said to be \defn{automatic}
    if 
    $S$ and $U_b$ (for each $b \in \Props$) are regular sets over some 
    non-empty finite alphabet 
    $\Sigma$, and each relation $\to_a$ (for each $a \in \ACT$) is given by a 
    transducer over $\Sigma$.
\end{definition}
We warn the reader that the most general notion of automatic transition systems 
\cite{BG04}, which allow non-length preserving transducers, are not needed
in this paper.
When the meaning is understood, we shall confuse the notation $\to_a$ for the
transition relation of $\struct$ and the transducer that recognises it.

\begin{example}
\label{ex:token}
\em
\OMIT{
Our first example is the Israeli-Jalfon probabilistic self-stabilising protocol
\cite{IJ90} (also see \cite{Norman04}).
The protocol has a 
ring topology, and each process either holds a token (denoted by $\top$) or
does not hold a token (denoted by $\bot$). At any given step, the Scheduler 
chooses a process $P$ that holds a token. The process $P$ can then pass the
token to its left or right neighbour each with probability 0.5. In doing so,
two tokens that are held by a process are merged into one token (held
by the same process).
}
We shall now model Israeli-Jalfon Protocol as an automatic transition system 
$\struct= \transysMDP$, 
where 
Scheduler's
actions are labeled by 1 and Process's actions are labeled by 2. 
In general, configurations of Israeli-Jalfon protocol are circular structures,
but they can easily be turned into a word over a certain finite alphabet by
linearising them.
More precisely, the domain $S$ of $\struct$ is the set of words over
$\ialphabet = \{\bot,\top,\hat{\top}\}$ of the form 
$(\bot+\top)^*\top(\bot+\top)^*$, or $(\bot+\top)^*\hat{\top}(\bot+\top)^*$.
\OMIT{
\medskip
\noindent
\begin{minipage}[t]{0.45\linewidth}
\begin{itemize}
    \item $(\bot+\top)^*\top(\bot+\top)^*$, or
\end{itemize}
\end{minipage}
\hfill
\begin{minipage}[t]{0.5\linewidth}
\begin{itemize}
    \item $(\bot+\top)^*\hat{\top}(\bot+\top)^*$.
\end{itemize}
\end{minipage}

\medskip
\noindent
}

For example, the configuration $\top\bot\top\bot$ denotes the configuration
where the 1st and the 3rd (resp. 2nd and 4th) processes are (resp. are not)
holding a token.
The letter $\hat{\top}$ is used to denote that Scheduler chooses a specific
process that holds a token.
Note that the intersection of languages generated by these two regular 
expressions is empty. The transition relation $\to_1$ is given by the 
regular expression
$
    I^* \biword{\top}{\hat{\top}} I^*
$
where $I := \{\biword{\top}{\top},\biword{\bot}{\bot}\}$.  
The transition relation $\top_2$ is given by a 
union of the following regular expressions:

\medskip
\noindent
\begin{minipage}[t]{0.45\linewidth}
\begin{itemize}
    \item $I^*
        \biword{\hat{\top}}{\bot}\left(\biword{\bot}{\top}+
        \biword{\top}{\top}\right) I^*$
    \item $I^* \left(\biword{\bot}{\top}+\biword{\top}{\top}\right)
        \biword{\hat{\top}}{\bot} I^*$
\end{itemize}
\end{minipage}
\hfill
\begin{minipage}[t]{0.5\linewidth}
\begin{itemize}
    \item $\left(\biword{\bot}{\top}+\biword{\top}{\top}\right)
        I^*\biword{\hat{\top}}{\bot}$
    \item
        $\biword{\hat{\top}}{\bot}I^*\left(\biword{\bot}{\top}+
        \biword{\top}{\top}\right)$
\end{itemize}
\end{minipage}

\medskip
\noindent
Note that the right column represents transitions that handle the circular
case. Also, note that if $I_0 = (\bot + \top)^*\top(\bot + \top)^*$ and
$F = \bot^*\top\bot^*$, Player 2 can always win the game from any
reachable configuration (note: $post_{\to^*}(I_0) = I_0$) by simply 
minimising the distance between the leftmost token and the rightmost token
in the configuration.
\qed
\end{example}

\subsection{Algorithm for liveness (an overview)}
\label{sec:progress}
\OMIT{
In the previous section, we have reformulated liveness over probabilistic
parameterised systems as a non-stochastic game over parameterised systems
(cf. Proposition \ref{prop:removeProb}). The resulting decision problem
(Item 2 of Proposition \ref{prop:removeProb}) can be stated as follows:
}
Our discussion thus far has led to a reformulation of liveness for 
probabilistic parameterised systems as the following decision problem:
given an automatic arena $\struct = \transysMDP$, a regular set 
$I_0 \subseteq S$ of initial configurations, and a regular set $F$ of final 
configurations, 
decide if Player 2 can force the game to reach $F$ in $\struct$ starting from 
each configuration in $post_{\rightarrow^*}(I_0)$. 
In the sequel, we will call $\langle \struct, I_0, F\rangle$ a \defn{game
instance}.
Note that the aformentioned problem is undecidable
even when $\to_2$ is restricted to identity relations, which amounts to the 
undecidable problem of safety \cite{rmc-survey}.
We will show now that decidability can be retained if ``advice bits'' are 
provided in the input. 


\defn{Advice bits} are a pair $\langle A, \prec\rangle$, 
where
$A \subseteq S$ is a set of game configurations and $\prec\ \subseteq S 
\times S$ is a binary relation over the game configurations.
Intuitively, $A$ is an inductive invariant, whereas $\prec$
is a well-founded relation that guides Player 2 to win.
More precisely,
the advice bits $\langle A, \prec \rangle$ are said to \emph{conform} to
the game instance $\langle \struct, I_0, F \rangle$ if:
        \begin{description}
        \item[(L1)] $I_0 \subseteq A$,
        \item[(L2)] $A$ is $\to$-inductive, i.e., 
            $
                \forall x, y: x \in A \wedge (x \to y) \Rightarrow y \in A
            $,
        \item[(L3)] $\prec\ $ is a \emph{strict preorder}\footnote{
A binary 
relation $\prec$ on a set $A$ is said to be a \defn{strict preorder} if it 
is irreflexive (i.e. for each $s \in A$, $s \not\prec s$) and transitive (for 
each $s,s',s'' \in A$, $s \prec s'$ and $s' \prec s''$ implies that 
$s \prec s''$). 
} on $S$, 
        \item[(L4)] Player 2 can progress from $A$ by following 
            $\prec$:
            \begin{equation*}
              \forall x \in A\setminus F, y \in S\setminus F:\; \big(
              (x \to_1 y) ~\Rightarrow~
              (\exists z \in A:\;
              (y \to_2 z) \wedge x \succ z)
              \big)~.
            \end{equation*}
    \end{description}
    Conditions \textbf{(L1)} and \textbf{(L2)} ensure
    that $post_{\to^*}(I_0) \subseteq A$, while conditions
    \textbf{(L3)}--\textbf{(L4)} ensure that Player 2 has a winning
    strategy from each configuration in $post_{\to^*}(I_0)$. Note that
    \textbf{(L3)} implies
    well-foundedness of $\prec$, provided that $\prec$
    only relates words of the same length (which is always sufficient for
    advice bits, and will later follow from the
    use of length-preserving transducers to represent $\prec$).





\begin{theorem}
    \label{thm:reachabilityCond}
    Let $\struct = \transysMDP$ be a $\to^*$-image-finite arena, i.e., 
    $post_{\to^*}(s)$ is finite, for each $s \in S$. Given a set $I_0 \subseteq 
    V_1$ of 
    initial configurations, and a set $F \subseteq V_1$ of final configurations,
    the following are equivalent:
    \begin{enumerate}
    \item Player 2 has a winning strategy reaching $F$ in $\struct$ 
        starting from each configuration in $post_{\rightarrow^*}(I_0)
        \cap V_1$. 
    \item There exist advice bits $\langle A, \prec\rangle$ conforming
        to the input $\langle \struct, I_0, F \rangle$.
      \OMIT{
        two sets $A \subseteq B$ and a binary relation $\prec\ 
        \subseteq S \times S$ such that
        \begin{description}
        \item[(L0)] $I_0 \subseteq A$,
        \item[(L1)] $A$ is $\to$-inductive, i.e., 
            \[
                \forall x, y: x \in A \wedge x \to y \Rightarrow y \in A
            \]
        \item[(L2)] $\prec\ $ is a strict preorder on $S$,
        \item[(L3)] Player 2 can force a win from $B$ following $\prec$:\\
            $\forall x \in B, y \in S\setminus F:\; \left(
        \begin{array}{@{}l@{}}
        x \to_1 y ~\Rightarrow~
        \\
            \exists z \in B \cup F:\;
        y \to_2 z \wedge x \succ z
        \end{array}
       \right)$.
    \end{description}
}
    \end{enumerate}
\end{theorem}
Advice bits $\langle A, \prec\rangle$ are said to be \defn{regular} if $A$
(resp. $\prec$) is given as a regular set (resp.~relation). 
With the help of regular advice bits, the problem of deciding a winning
strategy for Player 2 becomes decidable:
\begin{lemma}
    Given an automatic arena $\struct = \transysMDP$, a regular set 
    $I_0 \subseteq S$ of initial configurations, a regular set $F$ of final 
    configurations, and regular advice bits $T = \langle A, 
    \prec\rangle$, we can effectively decide whether $T$ conforms to
    the game instance $\langle \struct,I_0,F \rangle$.

    \label{lm:advice}
\end{lemma}
Lemma \ref{lm:advice} follows from the fact that each of the conditions 
\textbf{(L1)}--\textbf{(L4)} is
expressible in first-order logic interpreted over the given game instance
extended with the advice bits, i.e., the transition systems
$\langle S; \mbox{$\{\to_1,\to_2,\prec\}$}, \{I_0,F,A\}\rangle$. 
Decidability then follows since model checking first-order logic formulas
over automatic transition systems is decidable (e.g. see \cite{Blum99,BG04}
and see \cite{anthony-thesis} for a detailed complexity analysis), the proof of which
is done by standard automata methods.

To decide whether Player 2 has a winning strategy for the reachability game,
Lemma~\ref{lm:advice} tells us that
one can systematically enumerate all possible regular advice bits and check
whether they conform to the input game instance $\langle \struct, I_0, 
F\rangle$. A naive enumeration would simply go through each $k=1,2,\ldots$
and all advice bits 
$\langle A, \prec\rangle$ where each of the two automata have at most
$k$ states. This would be extremely slow. 

%% file: autoReachability.tex
\section{Automatic liveness proofs}
\label{sec:autoReachability}

We now describe how regular advice bits~$\langle A, \prec\rangle$ for
(regular) game instances~$\langle \struct, I_0, F \rangle$ can be
computed automatically, thus proving that Player~2 can win from every
reachable configuration, 
which (as we saw in the previous section) establishes liveness for
randomised parameterised systems. We define a constraint-based method
that derives $\langle A, \prec\rangle$ as the solution of a set of
Boolean formulas representing the
conditions~\textbf{(L1)}--\textbf{(L4)} from
Section~\ref{sec:progress}. Since a full Boolean encoding of
\textbf{(L1)}--\textbf{(L4)} would be exponential in the size of the
automata representing the advice bits, 
  our algorithm starts with a relaxed version of
\textbf{(L1)}--\textbf{(L4)} and gradually refines the encoding with
the help of counterexamples; in this sense, our approach is an
instance of CEGAR~\cite{DBLP:conf/cav/ClarkeGJLV00}, and has
similarities with recent learning-based methods for computing
inductive invariants \cite{Neider13}.

Throughout the section we assume that an alphabet~$\ialphabet$ and game
instance~$\langle \struct, I_0, F \rangle$ has been fixed.  We will
represent the well-founded relation~$\prec$ using a
transducer~$\TraPrec = (\ialphabet \times
\ialphabet,\controls_\prec,\delta_\prec,q^0_\prec,F_\prec)$, and the
set~$A$ as automaton~$\Aut_A =
(\ialphabet,\controls_A,\delta_A,q^0_A,F_A)$.  Our overall approach
for computing the automata makes use of two main components, which are
invoked iteratively within a refinement loop:
\begin{description}
\item[\textsc{Synthesise}] Candidate automata~$(\AutABA, \TraPrec)$
  with $n_A$ and  $n_\prec$ states, respectively, are computed
  simultaneously with the help of a SAT-solver, enforcing a relaxed
  set of conditions encoded as a Boolean constraint~$\psi$. The
  transducer~$\TraPrec$ is length-preserving and irreflexive by
  construction; this implies that the relation~$\prec$ is a
  well-founded preorder iff it is transitive.
\item[\textsc{Verify}] It is checked whether the automata~$(\AutABA,
  \TraPrec)$ satisfy conditions~\textbf{(L1)}--\textbf{(L4)}
  from Section~\ref{sec:progress}. If this is not the case, $\psi$ is
  strengthened to eliminate counterexamples, and \textsc{Synthesise}
  is again invoked; otherwise, $(\AutABA, \TraPrec)$
  represent a winning strategy for Player~2 by
  Theorem~\ref{thm:reachabilityCond}.
\end{description}

This refinement loop is enclosed by an outer loop that increments the
parameters~$n_A$, and $n_\prec$ (initially set to some small
number) when \textsc{Synthesise} determines that no automata
satisfying $\psi$ exist anymore. Initially, the formula~$\psi$
approximates \textbf{(L1)}--\textbf{(L4)}, by capturing aspects that
can be enforced by a Boolean formula of polynomial size.  The next
sections described \textsc{Synthesise} and \textsc{Verify} in detail.


\subsection{\textsc{Verify}: checking \textbf{(L1)}--\textbf{(L4)}
precisely}
\label{sec:verify}

Suppose that automata~$(\AutABA, \TraPrec)$ have been computed.  In
the \textsc{Verify} stage, it is determined whether the automata
indeed satisfy the conditions \textbf{(L1)}--\textbf{(L4)}, which can
effectively be done due to Lemma~\ref{lm:advice}. The check will have
one of the following outcomes:
\begin{enumerate}
\item $(\AutABA, \TraPrec)$ represent correct advice bits.
\item \textbf{(L1)} is violated: some word~$x \in I_0$ is not accepted
  by $\AutABA$.
\item \textbf{(L2)} is violated: there are words~$x \in A$ and $y$ with
  $x \to y$, but $y \not\in A$.
\item \textbf{(L3)} is violated: $\TraPrec$ does not represent a
  transitive relation (recall that $\TraPrec$ is length-preserving and
  irreflexive by construction).
\item \textbf{(L4)} is violated: there are words~$x \in A\setminus F$
  and $y \in S \setminus F$ such that $x \to_1 y$, but no word~$z \in
  A$ exists with $y \to_2 z$ and $x \succ z$.
\end{enumerate}
In cases 2--5, the computed words are counterexamples that are fed
back to the \textsc{Synthesise} stage; details for this are given in
Sect.~\ref{sec:counterexamples}.

\OMIT{
\medskip
We will use techniques from automatic structures \cite{Blum99,BG04} to
verify \textbf{(L1)}--\textbf{(L4)}, i.e., expressing them as
first-order formulas over the vocabulary $\sigma$ consisting of all
binary relations recognised by transducers and all regular languages
(represented as NFA). To make this more precise, we will use the
following folklore result (e.g., \cite{anthony-thesis}):
\begin{proposition}
  Given a quantifier-free first-order logic formula $\varphi$ over the
  vocabulary $\sigma$ of the form
  \begin{equation}
    \label{eq:automataFor}
    C_1 \wedge \cdots \wedge C_n,
  \end{equation}
  where each $C_i$ is a clause over $\sigma$ (i.e. a disjunction of literals
  over propositions of the form $R(x,y)$ and $L(x)$, where $R$ is a transducer
  and $L$ is a regular language), we can construct an automaton equivalent
  to $\varphi$
  in time exponential in $|\varphi|$. If $\varphi$ has no negative literal,
  then the construction is polynomial in $|\varphi|$ and exponential in
  $n$.
  \label{prop:folklore}
\end{proposition}
Notice that when $n$ is fixed (e.g. $n \leq 4$), then the construction
is polynomial when $\varphi$ has no negative literal. The proof of the
above proposition is standard (e.g. see \cite{anthony-thesis}):
conjunctions are handled by a product automata construction, while
negations are handled by complementing automata (hence, exponential
for NFA).

\medskip
}

The required checks on $(\AutABA, \TraPrec)$ can be encoded as
validity of first-order formulas, and finally carried out using
automata methods (e.g. see \cite{anthony-thesis}). In \textbf{(L3)}
and \textbf{(L4)}, it is in addition necessary to eliminate the
quantifier~$\exists z$ by means of projection. Note that all free
variables in the formulas are implicitly universally quantified.
\begin{center}
  \begin{tabular}{l@{\quad}l}
    \textbf{(L1)}
    &
    $I_0(x) \Rightarrow A(x)$
    \\[0.5ex]
    \textbf{(L2)}
    &
    $A(x) \wedge (x \to_1 y \vee x \to_2 y) \Rightarrow A(y)$
    \\[0.5ex]
    \textbf{(L3)}
    &
    $x \prec y \wedge y \prec z \Rightarrow x \prec z$
    \\[0.5ex]
    \textbf{(L4)}
    &
    $A(x) \wedge \neg F(x) \wedge \neg F(y) \wedge (x \to_1 y) \Rightarrow
    \exists z.\; \big(
    A(z) \wedge (y \to_2 z) \wedge x \succ z
    \big)$
  \end{tabular}
\end{center}


\subsection{\textsc{Synthesise}: computation of candidate automata}
\label{sec:guessing}

We now present the Boolean encoding used to search for (deterministic)
automata~$(\AutABA, \TraPrec)$, and to this end make the simplifying
assumption that the states of the transducer~$\TraPrec$ are
$\controls_\prec = \{1,\ldots,n_\prec\}$, states of the
automaton~$\AutABA$ are $\controls_A = \{1,\ldots,n_A\}$, and that
$q^0_\prec = q^0_A = 1$ are the initial states.  The following Boolean
variables are used to represent automata: a variable~$x^\prec_{t}$ for
each tuple $t = (q,a,b,q') \in \controls_\prec \times \ialphabet
\times \ialphabet \times \controls_\prec$; a variable~$x^A_{t}$ for
each tuple $t = (q,a,q') \in \controls_A \times \ialphabet \times
\controls_A$; and a variable~$z^M_q$ for each $q \in \controls_M$ and
$M \in \{\prec, A\}$.
The assignment~$x^M_t = 1$ is interpreted as the existence of the
transition~$t$ in the automaton for $M$; likewise, we use $z^M_q = 1$
to represent that $q$ is an accepting state (in DFAs it is in general
necessary to have more than one accepting state).

The set of considered automata in step~\textsc{Synthesise} is
restricted by imposing a number of conditions. Most importantly, only
deterministic automata are considered, which is important for
refinement: to eliminate counterexamples, it will be necessary to
construct Boolean formulas that state \emph{non-acceptance} of certain
words, which can only be done succinctly in the case of languages
represented by DFAs:
\begin{description}
\item[(C1)] The automata~$\AutABA$ and $\TraPrec$ are deterministic.
\end{description}
The second condition encodes irreflexivity of the relation~$\prec$:
\begin{description}
\item[(C2)] Every accepting path in $\TraPrec$ contains a label~$(a,
  b)$ with $a \not= b$.
\end{description}
The third group of conditions captures minimality properties: automata
that can (obviously) be represented with a smaller number of states
are excluded:
\begin{description}
\item[(C3)] Every state of the automata~$\AutABA$ and $\TraPrec$ is
  reachable from the initial state.
\item[(C4)] From every state in the automata~$\AutABA$ and $\TraPrec$
  an accepting state can be reached.
\end{description}
Finally, we can observe that the states of the constructed automata
can be reordered almost arbitrarily, which increases the search space
that a SAT solver has to cover. The performance of \textsc{Synthesise}
can be improved by adding \emph{symmetry breaking}
constraints. Symmetries can be removed by asserting that automata
states are sorted according to some structural properties extracted
from the automaton; suitable properties include whether a state is
accepting, or which self-transitions a state has:
\begin{description}
\item[(C5)] The states~$\{2, \ldots, n_M\}$ (for $M \in \{\prec, A\}$)
  are sorted according to the integer value of the bit-vector
  $
    \langle z^M_q, x^M_{(q, l_1, q)}, \ldots, x^M_{(q, l_k, q)} \rangle
    $
  where $q \in \{2, \ldots, n_M\}$ and $l_1, \ldots, l_k$ is some
        fixed order of the \anthonychanged{transition} labels in $M$.
\end{description}
\anthony{Are you sure that the first and last arguments of $(q,l_i,q)$
  should be both $q$, not a new $q'$?}  \philipp{that should be
  correct. the idea is that we need to find some structural properties
  of the automata states that can be used for sorting; i.e.,
  properties that are independent of the numbering of states. here, we
  use whether a state is accepting, and the labels of self-transitions
  for sorting. I tried to improve the explanation before C5, better now?}

\paragraph{Encoding as formulas}

The encoding of \textbf{(C1)} and \textbf{(C5)} as a Boolean
constraint is straightforward.
\anthony{Are you sure encoding of \textbf{(C5)} as a formula is obvious? I 
think there are several ways of encoding this condition.}%
\philipp{there are several ways, but I think the details are not overly
important here? in the end, the encoding just boils down to lexicographic
comparison of bit-vectors} For \textbf{(C2)}, 
we assume additional
Boolean variables~$r_q$ (for each $q \in \controls_\prec$) to identify
states that can be reached via paths with only $(a, a)$
labels. \textbf{(C2)} is ensured by the following constraints, which
are instantiated for each $q \in \controls_\prec$:
\begin{equation*}
  (q \not= q^\prec_0) \vee r_q,
  \qquad
  \neg z^\prec_q \vee \neg r_q,
  \qquad
  \neg r_q \vee
  \bigwedge_{a \in \ialphabet, q' \in \controls_\prec}
  (\neg x^\prec_{(q, a, a, q')} \vee r_{q'}).
\end{equation*}
The first constraint ensures that $r_q$ holds for the initial state,
the second constraint excludes $r_q$ for all final states. The third
constraint expresses preservation of the $r_q$ flags under $(a, a)$
transitions.

\medskip
We outline further how \textbf{(C3)} can be encoded for $\AutABA$ (the
other parts of \textbf{(C3)} and \textbf{(C4)} are similar). We assume
additional variables~$y_q$ (for each $q \in \controls_A$) ranging over
the interval $[0,n_A-1]$, 
\anthony{This interval looks suspicious since 0 is not a state in $\controls_A$}
\philipp{but $y_q$ represents the distance from the initial state; this distance
can be 0?}
    to encode the distance of a state from the
initial state; these integer variables can further be encoded in
binary as a vector of Boolean variables. The following formulas,
instantiated for each $q \in \controls_A$, define the value of the
variables, and imply that every state is only finitely many
transitions away from the initial state:
\begin{equation*}
  y_1 = 0,
  \qquad
  (q = 1) \vee \bigvee_{a \in \ialphabet, q' \in \controls_A}
  \big(x^A_{(q', a, q)} \wedge y_q = y_{q'} + 1\big)~.
\end{equation*}

\subsection{Counterexample elimination}
\label{sec:counterexamples}

If the \textsc{Verify} step discovers that $(\AutABA, \TraPrec)$
violate some of the required conditions~\textbf{(L1)}--\textbf{(L4)},
one of four possible kinds of counterexample will be derived,
corresponding to outcomes~\#2--\#5 described in
Sect.~\ref{sec:verify}. The counterexamples are mapped to
constraints~$\mathit{CE}_i$ (for $i = 1, \ldots, 4$) to be added to
$\psi$ in \textsc{Synthesise} \anthonychanged{as a conjunct}:
\begin{itemize}
\item A configuration~$x$ from $I_0$ has to be included in $A$:
  ~~$\mathit{CE}_1 ~=~ A(x)$
\item A configuration~$y$ has to be included in $A$, under the
  assumption
  that $x$ is included:
  $\mathit{CE}_2 ~=~ \neg A(x) \vee A(y)$
\item Configurations~$x, z$ have to be related by $\prec$, under the
  assumption that $x, y$ and $y, z$ are related:
  ~~$\mathit{CE}_3 ~=~ x \not\prec y \vee y \not\prec z \vee x
  \prec z$
\item Player 2 has to be able to make a $\prec$-decreasing step from
  $y$, assuming $x \to_1 y$ and $x$ is included in $A$:
  ~~$\mathit{CE}_4 ~=~
  \neg A(x) \vee \exists z.\; \big( A(z) \wedge (y \to_2
  z) \wedge x \succ z \big)$
\end{itemize}
Each of the formulas can be directly translated to a Boolean
constraint over the vocabulary introduced in Sect.~\ref{sec:guessing},
augmented with additional auxiliary variables; the most intricate case
is $\mathit{CE}_4$, due to the quantifier~$\exists z$. More details
are given in \shortlong{the full version}{Appendix~\ref{app:cex}}.


\section{Optimisations and incremental liveness proofs}
\label{sec:optimisations}

The monolithic approach introduced so far is quite fast when compact
advice bits exist (as shown in Sect.~\ref{sec:experiments}), but
tends to be limited in scalability for more complex systems, because
the search space grows rapidly when increasing
the size of the considered automata. To address this issue, we
introduce a range of optimisations of the basic method, in particular
an \emph{incremental} algorithm for synthesising advice bits, computing
the set~$A$ and the relation~$\prec$ by repeatedly constructing
small automata.


\subsection{Incremental liveness proofs}
\label{sec:incremental}

We first introduce a disjunctive version of the advice bits used to
witness liveness:
\begin{definition}
    Let $(J, <)$ be a non-empty well-ordered \anthony{Do we need to
    add ``finite'' here?}\philipp{no, everything also works for infinite sets. there
  is in fact a comment further down in the text. we need infinite $J$
  to accelerate using symmetries} index 
    set.\footnote{This
    means, $<$ is a strict total well-founded order on $J$.} A
  \emph{disjunctive advice bit} is a tuple~$\langle A, (B_j, \prec_j)_{j \in
    J}\rangle$, where $A, B_j \subseteq S$ are sets of game configurations,
  and each $\prec_j\, \subseteq S \times S$ is a binary relation over
  the game configurations, such that:
  \begin{description}
  \item[(D1)] $I_0 \subseteq A$;
  \item[(D2)] $A$ is $\to$-inductive, i.e.,
    $
    \forall x, y: x \in A\setminus F \wedge (x \to y) \Rightarrow y \in A
    $;
  \item[(D3)] $A$ is covered by the $B_j$ sets and $F$, i.e.,
    $
    A \subseteq F \cup \bigcup_{j \in J} B_j
    $;
  \item[(D4)] for each $j \in J$, the relation~$\prec_j$ is a strict
    preorder on $S$;
  \item[(D5)] for each $j \in J$, player~2 can progress from $B_j$ by
    following $\prec_j$:
    \[
    \forall x \in A \cap B_j \setminus (F \cup \bigcup_{i < j} B_i),
    y \in S\setminus F:\; \left(
      \begin{array}{@{}l@{}}
        (x \to_1 y) ~\Rightarrow~
        \\
        \exists z \in B_j:\;
        (y \to_2 z) \wedge z \prec_j x
      \end{array}
    \right)~.
    \]
  \end{description}
\end{definition}

The difference to monolithic advice bits (as defined in
Sect.~\ref{sec:progress}) is that the global preorder~$\prec$ is
replaced by a set of preorders~$\prec_j$. Player~2 progresses to
sets~$B_i$ with smaller index~$i < j$ by following $\prec_j$, and this
way eventually reaches $F$. A monolithic order~$\prec$ can be
reconstructed by defining
\begin{equation*}
  x \prec y ~~\Leftrightarrow~~
  \begin{cases}
    \mathit{idx}(x) < \mathit{idx}(y) &
    \text{if~} \mathit{idx}(x) \not= \mathit{idx}(y)
    \\
    x \prec_j y &
    \text{if~} \mathit{idx}(x) = \mathit{idx}(y) = j
  \end{cases}
\end{equation*}
where $\mathit{idx}(x) = \min \{ j \in J \mid x \in B_j \}$, and
$\mathit{idx}(x) = \min J$ in case these is no $j\in J$ with $x \in
B_j$. From this, it immediately follows that
Theorem~\ref{thm:reachabilityCond} also holds for disjunctive advice
bits.  We can further note that if $J$ is finite and all sets in $(A,
(B_j, <_j)_{j \in J})$ are regular, then the disjunctive advice bits
correspond to regular monolithic advice bits; in general this is not
the case for infinite~$J$.

\begin{algorithm}[tb]
  $A \leftarrow S$ \atcp{ Over-approximation of reachable configurations}
  $W \leftarrow F$ \atcp{Under-approximation of winning configurations}

  \medskip
  \While{$A \not\subseteq W$}{
    choose a word~$u \in A \setminus W$\;
    \If{$u$ is reachable}{
      $W \leftarrow W \cup \mathit{win}(u, A, W)$
      \atcp{Widen set of winning configurations}
    }
    \Else{
      $A \leftarrow A \cap \mathit{invariant}(u, A)$
      \atcp{Tighten set of reachable configurations}
    }
  }

  \medskip
  \Return{``Player~2 can win from every reachable configuration!''}

  \caption{Incremental liveness checker}
  \label{alg:disjunctive}
\end{algorithm}

Alg.~\ref{alg:disjunctive} outlines the incremental liveness checker,
defined with the help of disjunctive advice bits. The algorithm
repeatedly refines a set~$A$ over-approximating the reachable
configurations, and a set~$F$ under-approximating the configurations
from which player~2 can win, and terminates as soon as all reachable
configurations are known to be winning. The algorithm makes use of two
sub-routines: in line~8, $\mathit{invariant}(u, A)$ denotes a
\emph{relatively inductive} invariant~$I$~\cite{Bradley2008} excluding
$u$, i.e., a set~$I \subseteq S$ such that
\begin{description}
\item[(RI1)] $u \not\in I$;
\item[(RI2)] $I_0 \subseteq I$;
\item[(RI3)] $A$ is $\to$-inductive relative to $A$,
  i.e., $ \forall x, y: x \in
  (I \cap A\setminus F) \wedge (x \to y) \Rightarrow y \in I $~.
\end{description}
If $A$ satisfies conditions~\textbf{(D1)} and \textbf{(D2)}, and $I$
is inductive relative to $A$, then also $A \cap I$ is an inductive set
in the sense of \textbf{(D1)} and \textbf{(D2)}. We can practically
compute automata representing sets~$I$ using a SAT-based refinement
loop similar to the one in Sect.~\ref{sec:autoReachability}.

The second function~$\mathit{win}(u, A, W)$ (line~6) computes a
further progress pair~$(B, \prec)$ witnessing the ability of Player~2
to win from $u$, and returns the set~$B$, subject to:
\begin{description}
\item[(PP1)] $u \in B$;
\item[(PP2)] the relation~$\prec$ is a strict preorder on $S$;
\item[(PP3)] Player~2 can progress from $B$ by following $\prec$:
    \[
    \forall x \in A \cap B \setminus W,
    y \in S\setminus F:\; \big(
      (x \to_1 y) ~\Rightarrow~
      \exists z \in B:\;
      (y \to_2 z) \wedge z \prec x
    \big)~.
    \]
\end{description}
Again, a SAT-based refinement loop similar to the one in
Sect.~\ref{sec:autoReachability} can be used to find regular progress
pairs~$(B, \prec)$ satisfying the conditions.  Comparing
\textbf{(RI1)}--\textbf{(RI3)} and \textbf{(PP1)}--\textbf{(PP3)} with
\textbf{(D1)}--\textbf{(D5)}, it is also clear that disjunctive advice
bits can be extracted from every successful run of
Alg.~\ref{alg:disjunctive}, which implies
soundness. Alg.~\ref{alg:disjunctive} is in addition complete in the
following sense: if there exist (monolithic) regular advice bits
conforming to a game~$\langle \struct, I_0, F \rangle$, if the
words~$u$ chosen in line~4 are always of minimum length, and if the
functions~$\mathit{invariant}$ and $\mathit{win}$ always compute
minimum-size automata (representing sets~$I$ and $(B, \prec)$) solving
the conditions \textbf{(RI1)}--\textbf{(RI3)} and
\textbf{(PP1)}--\textbf{(PP3)}, then Alg.~\ref{alg:disjunctive}
terminates. This minimality condition is satisfied for the
learning-based algorithms derived in Sect.~\ref{sec:autoReachability}.

\subsection{Pre-Computation of inductive invariants}
\label{sec:precomputedInvs}

Alg.~\ref{alg:disjunctive} can be optimised in different
regards. First of all, the assignment~$A \leftarrow S$ (line~1)
initialising the approximation~$A$ of reachable states can be replaced
with more precise pre-computation of the reachable states, for
instance with the help of abstract regular model
checking~\cite{BHV04}. In fact, any set~$A$ satisfying \textbf{(D1)}
and \textbf{(D2)} can be chosen.

We propose an efficient method for initialising $A$ by utilising
Angluin's $L*$-learning algorithm~\cite{Angluin:1987:LRS:36888.36889},
which is applicable due to the property of length-preserving arenas
that reachability of a given configuration $w$ (a word) from the initial
configurations~$I_0$ is \emph{decidable.} Decidability follows from
the fact that there are only finitely many configurations up to a
certain length, and the words occurring on a derivation~$w_0 \to w_1
\to \cdots \to w_n$ all have the same length, so that known
(explicit-state or symbolic) model checking methods can be used to
decide reachability. 

Reachability of configurations enables us to construct an $L*$
teacher (a.k.a. oracle). Membership queries for individual words~$w$ are 
answered by checking reachability of $w$ in the game. Once the learner
produces an hypothesis automaton~$\cal H$, the teacher verifies that:
\begin{enumerate}
\item $\cal H$ includes the language~$I_0$, i.e., \textbf{(D1)} is
  satisfied. If this is not the case, the teacher informs the learner
  about some further word in $I_0$ that has to be accepted by $\cal
  H$.
\item $\cal H$ is inductive, i.e., satisfies condition~\textbf{(D2)},
  which can be checked by means of automata methods (as in
  Sect.~\ref{sec:autoReachability}). If \textbf{(D2)} is violated, the
  counterexample pair~$(x, y)$ is examined, and it is checked whether
  the configuration~$x$ is reachable. If $x$ is not reachable, the
  teacher gives a negative answer and demands that $x$ be removed from
  the language; otherwise, the teacher demands that $y$ is added to
  the language.
\item $\cal H$ describes the precise set of reachable configurations,
  for configuration length up to some fixed $n$. In other words,
  whenever $\cal H$ accepts some word~$w$ with $|w| \leq n$, the
  configuration~$w$ has to be reachable; otherwise, the teacher
  demands that $w$ is eliminated from the language. \anthony{This
    sentence sounds weird. Are we supposed to remove the first ``not''
    and replace ``any'' but ``some''.}\philipp{rephrased, better?}
\end{enumerate}
\anthony{The last item sounds rather confusing. I think we need to say that
for each $n$, there is a teacher $T_n$. This teacher does the last check
up to words of length $n$. Also, in practice, we need to say that $n$ is
in the implementation (or put a reference to next Sec.). 
}
If all three tests succeed, the teacher accepts the produced
automaton~$\cal H$, which indeed represents a set~$A$ satisfying
\textbf{(D1)} and \textbf{(D2)}. Tests~1 and 2
ensure that $\cal H$ is an inductive invariant, while test~3 is
necessary to prevent trivial solutions: without  the  test, the algorithm
could always return an automaton~$\cal H$ recognising the universal
language~$\Sigma^*$. The parameter~$n$ determines the precision
of synthesised invariants: larger $n$ lead to automata~$\cal H$ that are
tighter over-approximations
of the precise language of reachable configurations.\footnote{In our
implementation we currently hard-code $n$ to be $5$.}\philipp{better?}

This algorithm is guaranteed to terminate if the set of reachable
configurations in an arena is regular; but it might only produce some
inductive over-approximation of the reachable configurations. In our
experiments, the computed languages usually capture reachable
configurations very precisely, and the learning process converges
quickly.


\subsection{Exploitation of game symmetries}
\label{sec:symmetries}

As a second optimisation, the incremental procedure can be improved to
take symmetries of game instances into account, thus reducing the
number of iterations needed in the incremental procedure; algorithms
to automatically find symmetries in parameterised systems have
recently proposed in \cite{DBLP:conf/vmcai/LinNR016}. This corresponds
to replacing line~6 of Alg.~\ref{alg:disjunctive} with the assignment
$
  W \leftarrow W \cup \sigma^*(\mathit{win}(u, A, W)) ;
$
where $\sigma$ is an \emph{automorphism} of the game instance~$\langle
\struct, I_0, F \rangle$, and $\sigma^*(L) = L \cup \sigma(L) \cup
\sigma^2(L) \cup \cdots$ represents unbounded application of $\sigma$
to a language~$L \subseteq \ialphabet^*$. An automorphism (or
\emph{symmetry pattern}~\cite{DBLP:conf/vmcai/LinNR016}) is a
length-preserving bijection~$\sigma : \ialphabet^* \to \ialphabet^*$
such that 1. initial and winning configurations are
$\sigma$-invariant, i.e., $\sigma(I_0) = I_0$ and $\sigma(F) = F$; and
2. $\sigma$ is a homomorphism of the moves, i.e., $u \to_i v$
if and only if $\sigma(u) \to_i \sigma(v)$ for $i \in \{1, 2\}$.

A symmetry commonly present in systems with ring topology is
\emph{rotation,} defined by $\sigma_{\text{rot}}(u_1u_2 \ldots u_n) =
u_2 \ldots u_n u_1$; the Israeli-Jalfon protocol
(Example~\ref{ex:token}) exhibits this symmetry, as do many other
examples. In addition, the
fixed-point~$\sigma_{\text{rot}}^*(L)$ can effectively be constructed
for any regular language~$L \subseteq \ialphabet^*$ using simple
automata methods, which is of course important for implementing the
optimised incremental algorithm.

In terms of disjunctive advice bits~$\langle A, (B_j, \prec_j)_{j \in
  J}\rangle$, application of a symmetry~$\sigma$ corresponds to
including a sequence $ (B, \prec), (\sigma(B), \prec^\sigma),
(\sigma^2(B), \prec^{\sigma^2}), \ldots $ of progress pairs, defining
$(u \prec^\rho v) \Leftrightarrow (\rho^{-1}(u) \prec \rho^{-1}(v))$
for any bijection~$\rho : \ialphabet^* \to \ialphabet^*$. 
\anthony{It took me a few minutes to parse $u\, \rho(\prec)\, v$. Any chance
we can rewrite it with something more intelligible?}\philipp{agreed; is
$u \prec^\rho v$ better?} The
resulting monolithic progress relation will in general not be regular;
in terms of ordinals, this means that a well-order~$(J, <)$ greater
than $\omega$ is chosen.


%% file: experiments.tex
\section{Experiments and Conclusion}
\label{sec:experiments}

\begin{table}[tb]
  \caption{Verification results for parameterised systems and
    games. \textbf{Mono} is the monolithic method from
    Sect.~\ref{sec:autoReachability}, \textbf{Incr} the incremental
    algorithm from Sect.~\ref{sec:incremental}, and \textbf{Inv} and
    \textbf{Symm} the optimisations introduced in
    Sect.~\ref{sec:precomputedInvs} and \ref{sec:symmetries},
    respectively. A dash~--- indicates that a model is not symmetric
    under rotation, or that the incremental algorithm is not applicable
    (in case of Take-away and Nim).
    The numbers in the table give runtime (wall-clock time)
    for the individual benchmarks and configurations; all experiments
    were done on an AMD Opteron 6282 32-core machine,
    Java heap memory limited to 20GB, timeout 2~hours.}
  \label{tab:experiments}
  
  \centering
    \begin{tabular}{@{~~~}l*{5}{@{\quad}c}@{~~~}}
      \hline
      & \textbf{Mono} & \textbf{Incr} &
      \textbf{Incr+Inv} &
      \textbf{Incr+Symm} &
      \textbf{Incr+Inv+Symm}
      \\\hline
      \multicolumn{6}{l}{\itshape Randomised parameterised systems}
      \\\hline
      Lehmann-Rabin (DP) \cite{DFP04} & T/O & T/O & T/O & 48min & 10min
      \\
      Israeli-Jalfon \cite{IJ90} & 4.6s & 22.7s & 21.4s & 9.9s & 9.7s
      \\
      Herman \cite{Her90} & 1.5s & 1.6s & 2.4s & --- & ---
      \\
      Firewire \cite{MMH03,EGK12} & 1.3s & 1.3s & 2.0s & --- & ---
      \\\hline
      \multicolumn{6}{l}{\itshape Deterministic parameterised systems}
      \\\hline
      Szymanski \cite{RMC-without,rmc-thesis} & 5.7s & 27min & 10min & --- & ---
      \\
      DP, left-right strategy & 1.9s & 6.4s & 3.4s & --- & ---
      \\
      Bakery \cite{RMC-without,rmc-thesis} & 1.6s & 2.7s & 1.9s & --- & ---
      \\
      Resource allocator \cite{donaldson-thesis} & 2.2s & 2.2s & 2.0s & --- & ---
      \\\hline
      \multicolumn{6}{l}{\itshape Games on infinite graphs}
      \\\hline
      Take-away \cite{Ferguson} & 2.8s & --- & --- & --- & ---
      \\
      Nim \cite{Ferguson} & 5.3s & --- & --- & --- & ---
      \\\hline
    \end{tabular}
\end{table}

%
%

All techniques introduced in this paper have been implemented in the
liveness checker \texttt{SLRP} \cite{SLRP} for parameterised systems, using the
SAT4J~\cite{DBLP:journals/jsat/BerreP10} solver for Boolean
constraints. 
For
evaluation, we consider a range of (randomised and deterministic)
parameterised systems, as well as Take-away and Nim games, shown in
Table~\ref{tab:experiments}. Two of the randomised protocols,
Lehmann-Rabin and Israeli-Jalfon are symmetric under rotation. 
Since Herman's original protocol in a ring \cite{Her90} only satisfies
liveness under ``fair'' schedulers, we used the version of the protocol in
a line topology, which does satisfy liveness under all schedulers. Firewire
is an example taken from \cite{EGK12,MMH03} representing a fragment of 
Firewire symmetry breaking protocol. For
handling combinatorial games, the monolithic method in
Sect.~\ref{sec:autoReachability} was adapted by removing
condition~\textbf{(L2)}; adaptation of
the incremental algorithm from Sect.~\ref{sec:incremental} to
this setting has not been considered yet.

\enlargethispage{2ex}
All models could be solved using at least one of the considered
\anthonychanged{CEGAR modes}. In most cases, the monolithic approach from
Sect.~\ref{sec:autoReachability} displays good performance, and in
case of the deterministic systems is competitive with existing tools
(e.g. \cite{AJRS06,rmc-thesis}). Monolithic reasoning outperforms the
incremental methods (Sect.~\ref{sec:optimisations}) in particular for
Szymanski, which is because Alg.~\ref{alg:disjunctive} spends a lot of
time computing a good approximation~$A$ of reachable states, although
liveness can even be shown using $A = \ialphabet^*$.

In contrast, the most complex model, the Lehmann-Rabin protocol for
Dining Philosophers, can only be solved using the incremental
algorithm, and only when accelerating the procedure by exploiting the
rotation symmetry of the game (Sect.~\ref{sec:symmetries}). In
configuration Incr+Inv+Symm, Alg.~\ref{alg:disjunctive} computes an
initial set~$A$ represented by a DFA with 23~states
(Sect.~\ref{sec:precomputedInvs}), calls the function~$\mathit{win}$
25~times to obtain further progress relations
(Sect.~\ref{sec:incremental}), and overall needs 4324~iterations of
the refinement procedure of Sect.~\ref{sec:autoReachability}.  To the
best of our knowledge, this is the first time that liveness under arbitrary
schedulers for
randomised parameterised systems like Lehmann-Rabin could be shown
fully automatically.


%% file: conc.tex
\medskip
\noindent
\textbf{Future Work}
We conclude with two concrete research questions among many others.
The most immediate question is how to embed fairness in
our framework of randomised parameterised systems. 
Another research direction concerns how to extend transducers to deal with data 
so as to model protocols where tokens may store arbitrary process IDs
(examples of which include Dijkstra's Self-Stabilizing Protocol
\cite{Dijkstra74}).


%% file: app.tex
\input{games}

\input{app-framework}
\input{app-cex}

%% file: games.tex
\section{Liveness as non-stochastic 2-player games}
\label{sec:games}

In this section, we shall justify our definition of liveness for 
randomised parameterised systems in terms of 2-player reachability games.
%
We will first review the necessary mathematical background.

\subsection{Markov Chains}
Before reviewing the definition of Markov Decision Processes, we will quickly
recall the definition of Markov chains (see \cite{marta-survey} for more
details).
A (discrete-time) \defn{Markov chain} (a.k.a. \defn{DTMC}) 
is a transition
system $\struct = \transysR$ equipped with a \defn{transition probability} 
function $\delta: R \to
(0,1]_{\mathbb{R}}$ such that $\sum_{q \in post(p)} \delta(q) = 1$ for each
$p \in S$. That is, $\delta$ associates each transition with its probability
of firing. 
Given a finite path $\ModelRun = s_0, \cdots, s_n$ from the initial state 
$s_0 \in S$, let $Run_{\ModelRun}$ be the set of all finite/infinite paths with
$\ModelRun$ as a prefix, i.e., of the form $\ModelRun \odot \ModelRun'$ for 
some finite/infinite
path $\ModelRun'$. Given a set $F \subseteq S$ of
target states, the probability $\Prob_{\struct}(s_0 \models \diamond F)$ 
(the subscript $\struct$ may be omitted when understood)
of reaching $F$ 
from $s_0$ in $\struct$ can be defined using a standard cylinder construction 
(e.g \cite{marta-survey}). That is, for each finite path 
$\pi = s_0, \cdots, s_n$ 
in $\struct$ from $s_0$, we set $Run_{\ModelRun}$ to be a basic cylinder,
to which we associate the probability $\Prob(Run_{\ModelRun}) =
\prod_{i=0}^{n-1} \delta((s_i,s_{i+1}))$.
This gives rise to a unique probability measure for the $\sigma$-algebra
over the set of all runs from $s_0$. The probability 
$\Prob(s_0 \models \diamond F)$ is then the probability of the event
containing the set of all paths with some ``accepting'' finite path as
a prefix, i.e., a finite path from $s_0$ ending in some state in $F$.

\subsection{Markov Decision Processes}
We recall some basic concepts on Markov decision processes (a.k.a. 
\defn{concurrent Markov chains}, e.g., see \cite{Var85,CY95}), especially how
liveness is defined over the model. 

A \defn{Markov decision process}
(\defn{MDP}) is a 
strictly alternating arena $\struct = \transysMDP$ such that $\langle S; 
\to_2\rangle$ 
is a DTMC (with some transition probability $\delta$).
Intuitively, the transition relation $\to_1$ is nondeterministic
(controlled by Scheduler), whereas the transition relation $\to_2$ is 
probabilistic. By definition of arenas, the configurations of the MDPs are
partitioned into the set $V_1$ of \emph{nondeterministic states} (controlled by 
Scheduler) and the set $V_2$ of \emph{probabilistic states}. In symbol, we
have
$pre_{\to_1}(S) \cap pre_{\to_2}(S) = \emptyset$.
Each Scheduler's strategy\footnote{Also called ``scheduler'' or 
``adversary'' for short.} $f: S.V_1 \rightarrow S$ \philipp{confusing, why
$SV_1$ and not $V_1$? \anthony{by definition, schedulers need not be memoryless.
But it will be mentioned below that memoryless restriction is ok.}} gives rise to an 
infinite-state DTMC
$\struct_f = \langle S'; \to_3 \rangle$ with the transition probability
$\delta'$ defined as follows. Here, $S'$ is the set of all 
finite/infinite paths $\pi$ from $s_0$. For each state $s' \in S$ and each path 
$\pi$ from $s_0$ ending in some state $s \in S$,
%
we define $\pi \to_3 \pi s'$ iff:
(1) if $s$ is a nondeterministic state, then $f(\pi s) = s'$, and
(2) if $s$ is a probabilistic state, then $s \to_2 s'$. 
Intuitively, $\struct_f$ is an unfolding of the game arena $\struct$ (i.e. a 
disjoint union of trees) where branching only occurs on probabilistic states.
Transitions
$\pi \to_3 \pi s'$
satisfying 
Case (1) have the probability $\delta'((\pi,\pi s')) := 1$; otherwise, its 
probability is $\delta'((\pi,\pi s')) := \delta((s,s'))$. Since $\struct_f$
is a DTMC, the quantity $\Prob_{\struct_f}(s_0 \models \diamond S^*F)$ is 
well-defined. Loosely speaking, this is the probability of reaching $F$ from
$s_0$ in $\struct$ under the scheduler $f$. The probability
$\Prob_{\struct,\mathcal{C}}(s_0 
\models \diamond F)$ of reaching $F$ from $s_0$ in the MDP $\struct$ under a
class $\mathcal{C}$ of schedulers
is defined
to be the infimum of the set of all probabilities $\Prob_{\struct_f}(s_0 \models
\diamond S^*F)$ over all $f \in \mathcal{C}$. We will omit mention of
$\mathcal{C}$ when it denotes the class of all schedulers.

In this paper, we are only concerned with the following \defn{liveness problem
for MDP}: given an MDP $\struct = \transysMDP$, a set $I_0 \subseteq S$ of 
initial states, and a set $F \subseteq S$ of target states, determine whether 
$\Prob_{\struct}(s_0 \models \diamond F) = 1$ for every initial state
$s_0 \in I_0$. Note that this is equivalent to proving that
$\Prob_{\struct_f}(s_0 \models \diamond S^*F) = 1$ for every initial state
$s_0 \in I_0$, and every scheduler $f$. 
Such a problem (which has many other
names: probabilistic universality, almost-sure probabilistic reachability, and 
almost-sure liveness) is commonly studied
in the context of MDPs (e.g. see \cite{CY95,Var85,LS07,marta-survey}).
In the case of probabilistic
parameterised systems
 $\mathcal{F} = \{\struct_i\}_{i \in \N}$, where each $\struct_i$ is a
 finite-state MDP, we may view $\mathcal{F}$ as an infinite-state MDP defined by
the disjoint union of $\struct_i$ over all $i \in \N$.  In this way, proving
liveness for $\mathcal{F}$ simply means proving liveness for \emph{each}
instance $\struct_i$ in $\mathcal{F}$.
%
%

\begin{convention}
    As in Convention \ref{con:games}, we make similar simplification
    for MDPs: (1) initial and final configurations belong to Player 1
    (i.e. $I_0, F \subseteq V_1$), and (2) non-final configurations are no dead 
    ends.
\end{convention}

\subsection{Removing probability}
\OMIT{
Given an arena $\struct = \transysMDP$, we say that it is \emph{turn-based} if
a move made by a player does not take the game back to her configuration, i.e.,
$\pi_2(\to_i) \cap \pi_1(\to_i) = \emptyset$, for each $i \in \{1,2\}$.
}
It is long known that the liveness problem for finite-state MDPs $\struct = 
\transysMDP$ depends on the topology of the graph $\struct$, not on the
actual probability values in $\struct$ (e.g. \cite{CY95,Var85,Alfaro99,HSP83}).
In this paper, we use an equivalent formulation of the problem in terms of 
2-player non-stochastic reachability games over $\struct$ \emph{viewed purely 
as an 
arena}, i.e., 
$\to_1$ defines the possible moves of Player 1 and $\to_2$ defines the possible
moves of Player 2 (ignore the transition probability $\delta$ associated with
$\to_2$). Since each instance of a probabilistic parameterised system
$\mathcal{F}$ is
a finite system, the liveness problem for $\mathcal{F}$ can similarly
be reformulated in terms of 2-player non-stochastic reachability games over
$\mathcal{F}$ (viewed as an arena).

\begin{proposition}
Given a probabilistic parameterised system $\mathcal{F} = \{\struct_i\}_{i \in
\N}$, a set $I_0 \subseteq V_1$ of initial states, and a set $F \subseteq V_1$ 
of final states, the following are equivalent:
\begin{enumerate}
\item $\Prob_{\struct_f}(q_0 \models \diamond F) = 1$ for every $\struct
    \in \mathcal{F}$, for every initial state $q_0 \in I_0$, and every 
    scheduler $f$. 
\item From each configuration $s_0 \in post_{\rightarrow^*}(I_0)$, Player 2 has 
    a winning strategy reaching $F$ in $\mathcal{F}$ (viewed
    as an arena).
\end{enumerate}
\label{prop:removeProb}
\end{proposition}
\OMIT{
In turn,
to answer questions about liveness over probabilistic parameterised systems, we 
need not specify the precise probability of a transition, i.e., we only
care about whether it has a \emph{positive} or \emph{zero} probability
of firing. 
}
This result is standard in the study of MDPs (e.g. see \cite{CY95,Var85,HSP83}).
For completeness sake, we provide a proof next.
\input{app-games}

%% file: app-games.tex
\subsection{Proof of Proposition \ref{prop:removeProb}}
\label{sec:removProb}
Before proving this proposition, let us first recall that
it suffices to consider ``simple'' winning strategies for reachability
games. More precisely, given an arena $\struct = 
\transysMDP$, a strategy $f: S^* V_i$ for a Player
$i$ is said to be \defn{memoryless} if $f(vp) = f(p)$ for all $p \in V_i$
and $v \in S^*$, i.e., $f$ depends only on the current configuration, not 
on the history of runs.  For 2-player reachability games, it is well-known that
\emph{if a player has a winning strategy, then she has a memoryless winning
strategy} \cite[Proposition 2.21]{ALG-book}. For notational simplicity, 
we will denote a memoryless strategy by a function mapping $V_i$ to $S$.

\textbf{(1) $\Rightarrow$ (2)}. To prove this, assume that (1) holds but (2) does
not. This means that from some $s_0 \in post_{\rightarrow^*}(I_0)$ Player 1
has a memoryless winning strategy that avoids $F$ in $\mathcal{F}$. In fact,
$s_0$ must belong to a specific instance $\struct^n = \langle S_n; \to_1, \to_2
\rangle$ of the parameterised
system $\mathcal{F}$ and that $s_0 \in post_{\rightarrow^*}(p_0)$, for some
$p_0 \in I_0 \cap S_n$.
Let $f: S_n^*V_1 \to S_n$ denote the aforementioned strategy of Player 1,
where $V_1 = pre_{\to_1}(S_n)$.
Since $s_0 \in post_{\rightarrow^*}(p_0)$,
there must exist a path
$\pi := p_0 \to_1 p_1 \to_2 \cdots \to_1 p_{m-1} \to_2 p_m$ in $\struct^n$ with 
$p_m = s_0$. We are now going to construct a new strategy $g: S_n^*V_1 \to S_n$
as follows. Let $\Pi$ denote the set of all nonempty prefixes of the finite path
$\pi$, i.e., $\{p_0,p_0p_1,\ldots,p_0p_1\ldots p_m\}$. Then, define
\[
    g(\sigma) = \left\{ \begin{array}{cc}
                             p_0p_1\ldots p_{i+1} & \text{ if 
                                        $\sigma = p_0\ldots p_i \in \Pi$ with 
                                        $i < m$, } \\
                             f(\sigma)  & \text{ otherwise.}
                        \end{array}
                \right.
\]
Consider the DTMC $\struct^n_g$ induced by the MDP $\struct^n$ under the strategy 
$g$. Then, following the path $\Pi$ in $\struct^n_g$ gets us to a configuration
$p_0p_1\ldots p_m$ ($p_m = s_0$) from which any configuration in $S^*F$ can never 
be visited. This proves that $\Prob_{\struct^n_g}(s_0 \models \diamond S^*F)
< 1$ and so $\Prob_{\struct^n}(s_0 \models \diamond F) < 1$.
 This contradicts our assumption of (1). In conclusion, (2) must hold.

\textbf{(2) $\Rightarrow$ (1)} 
We assume that if (1) does not hold, then (2)
does not hold. So, assuming that (1) does not hold, there must exist a
scheduler $f$ such that $\Prob_{\struct_f}(q_0 \not\models \diamond F) > 0$.
The set of paths from $q_0$ that avoids $F$ in $\struct_f$ (i.e. satisfying the 
formula $\neg \diamond F$) is known to be measurable \cite{Var85} and so must
contain the set $\Pi$ of all paths that have some finite path $\pi = q_0, 
\ldots, q_n$ in $\struct_f$ as a prefix. Following the scheduler's strategy, 
Player 1 can win the game from $q_n \in post_{\to^*}(I_0)$ avoiding $F$ in 
$\struct$. This proves that (2) does not hold.

\OMIT{
for some $\struct \in \mathcal{F}$ and some initial state $q_0$ in $\struct$.
We will show that Player 1 has a winning strategy avoiding $F$ from some
$s_0 \in post_{\to^*}(I_0)$ in $\struct$ (viewed as an arena). 
}

%% file: app-framework.tex
\section{Other examples and Missing proofs from Section \ref{sec:framework}}

\subsection{Other examples}
Here we provide descriptions of several other protocols that we consider in the
benchmark. The descriptions of the other protocols can be found with the
tool \cite{SLRP}.

\begin{example}
    \em
    Another example is the Lehmann-Rabin protocol for the dining philosopher
    problem \cite{LR81} (also see \cite{Lynch-book}). In this example, $n$
    philosophers sit at a round table. In between two philosophers, a fork
    is placed on the table. The problem is to ensure that, under all
    possible schedulers, eventually one philosopher must eat. 
    It is known that there is no \emph{symmetric} solution (i.e. all 
    philosophers are completely identical) to the problem if the philosophers
    are completely deterministic (e.g. see \cite{Lynch-book}). Lehmann-Rabin
    protocols \cite{LR81} shows that a symmetric solution exists when the
    philosophers are probabilistic. 
    Here we present the version of
    the protocol where idle transitions in the philosopher's program
    are removed when chosen by the scheduler
    (see \cite{DFP04}). 
    The alphabet $S$ is 
    \[
        \{T,H,\LeftW,\RightW,\LeftS,\RightS,E\}.
    \]
    Intuitively, $T$ (resp. $H$) denotes \emph{thinking} (resp. \emph{hungry}).
    The letter $\LeftW$ (resp. $\RightW$) denotes \emph{waiting for the
    left (resp. right) fork}. The letter $\LeftS$ (resp. $\RightS$) denotes 
    \emph{waiting for the right (resp. left) fork with the left (resp. right)
    forked already in hand}. Finally, the letter $E$ denotes that the
    philosopher is \emph{eating}. The initial states are $T^3T^*$ (containing 
    at least three philosophers). 
    Let 
    \[
        \hat{S} = \{ \hat{a} : a \in S \}
     \]
     and let $\ialphabet = S \cup \hat{S}$. Let 
     \[
         \LeftHold = \{ \LeftS,\LeftD, E \}
     \]
     and
     \[
         \RightHold = \{ \RightS,\RightD, E \}.
     \]
     We write $\neg \LeftHold$ (resp. $\neg \RightHold$) to mean the
     complement $S \setminus \LeftHold$ (resp.  $S \setminus \RightHold$).
    Let $I = \{ (p,q) \in S \times
    S : p = q \}$. Let us define the binary relation $\leadsto$:
    \begin{enumerate}
        \item $\widehat{T} \leadsto H$
        \item $\widehat{H} \leadsto \LeftW$ and $\widehat{H} \leadsto \RightW$
        \item $A\widehat{\LeftW} \leadsto A\LeftS$, for each $A \in \neg 
                \RightHold$.
            \item $\widehat{\RightW} A \leadsto \RightS A$, for each $A \in \neg\LeftHold$
            \item $\widehat{\LeftS}A \leadsto EA$, for each $A \in \neg\LeftHold$
            \item $\widehat{\LeftS} A \leadsto \LeftD A$, for each $A \in \LeftHold$
            \item $A \widehat{\RightS} \leadsto AE$, for each $A \in \neg\RightHold$
            \item $A \widehat{\RightS} \leadsto A\RightD$, for each $A \in\RightHold$
            \item $\widehat{\LeftD} \leadsto H$ and $\widehat{\RightD} \leadsto H$
    \end{enumerate}
    We describe the transducer $\to_1$ by a union of the regular
    expressions:
    \begin{itemize}
        \item $I^*(A,\hat{A})I^*$ for each symbol $A \in S$ with
            $\hat{A} \leadsto X$ for some $X \in S^*$.
        \item $I^*(A,\hat{A})(B,B)I^*$ for symbols $A, B \in S$ with
            $\hat{A}B \leadsto X$ for some $X \in S^*$
        \item $(B,B)I^*(A,\hat{A})$ for symbols $A, B \in S$ with
            $\hat{A}B \leadsto X$ for some $X \in S^*$
        \item $I^*(A,A)(B,\hat{B})I^*$ for symbols $A, B \in S$ with
            $A\hat{B} \leadsto X$ for some $X \in S^*$
        \item $(B,\hat{B})I^*(A,A)$ for symbols $A, B \in S$ with
            $A\hat{B} \leadsto X$ for some $X \in S^*$
    \end{itemize}
    The transducer $\to_2$ is described as a union of the regular
    expressions: 
    \begin{itemize}
        \item \anthonychanged{$I^*(\hat{X},X')I^*$, for each $X,X' \in S$ with
            $X \leadsto X'$}
        \item \anthonychanged{$I^*(X,X')(Y,Y')I^*$, for each $X,X' \in S \cup \hat{S}$,
            and $Y,Y' \in S$ with $XY \leadsto X'Y'$}
        \item \anthonychanged{$(Y,Y')I^*(X,X')$, for each $X,X' \in S \cup \hat{S}$
            and $Y,Y' \in S$ with $XY \leadsto X'Y'$}
    \end{itemize}
    The set $F$ of final configurations is $S^*ES^*$ with at least one
    philosopher eats. Duflot \emph{et al.} \cite{DFP04} gave a highly 
    non-trivial proof that this protocol satisfies liveness with probability 1
    under arbitrary schedulers.

\end{example}

\subsection{Proof of Theorem \ref{thm:reachabilityCond}}
Before proving the theorem,
we refer the reader to the first paragraph of Section \ref{sec:removProb}
to review the notion of determinacy and memoryless winning strategies.

    ($\Leftarrow$) By \textbf{(L0)} and \textbf{(L1)}, it follows that
    $post_{\to^*}(I_0) \subseteq A$. We define a 
    strategy $g: S^*(S\setminus F) \to A$ of Player 2. By \textbf{(L4)}, 
    we have a relation $R \subseteq (A \setminus F) \times (S \setminus F)
    \times A$ such that $(v_1,v_2,v_1') \in R$ iff $v_1 \to_1 v_2$, 
    $v_2 \to_2 v_1'$, and $v_1 \succ v_1'$. From this relation we can define
    a partial function $g: Path \to A$, where $Path$ is the set of all 
    paths from $I_0$ to $S \setminus F$, as follows: if $\pi \in Path$ is of 
    the form $wv_1v_2$ for some $w \in S^*$, $v_1 \in A \setminus F$, and 
    $v_2 \in V_2$, then $g(\pi)$ is defined as any element $v_1'$ such that 
    $(v_1,v_2,v_1') \in R$.
    Note that this is well-defined by our assumption that $\struct$ is
    $\to^*$-image-finite, i.e., since this implies that $\to_1$ and $\to_2$
    are image-finite. We claim that $g$ is a winning strategy for Player 2
    from each initial configuration $s_0 \in post_{\to^*}(I_0) \cap V_1$.
    To show this,
    take any arbitrary strategy $f: S^*V_1 \to S$ for Player 1, and consider 
    the unique
    path $\sigma: s_0 \to_1 s_1 \to_2 \cdots$ from $s_0$ defined by $f$ and 
    $g$. 
    By Assumption \textbf{(A2)}, we may assume that $f(w \cdot v_1)$ is defined
    whenever $v_1 \notin F$. By Assumption \textbf{(A0)}, each 
    configuration $s_{2k+1}$ belongs to
    $V_2$ (which do not intersect with $F$ by \textbf{(A1)}). 
    \textbf{(L4)} implies that
    each configuration $s_{2k}$ belongs to the set $A$ or the set $F$.
    In fact, \textbf{(L4)} also implies that 
        $s_0 \succ s_2 \succ s_4 \succ \cdots.$
    Since $\struct$ is $\to^*$-image-finite and $\succ$ is a strict preorder,
    this sequence $\sigma$ is finite and ends in some configuration $s_{2r}$ 
    for some
    $r \in \N$. By Assumption \textbf{(A2)}, $s_{2r}$ has to be in $F$; for,
    if not, then $s_{2r} \to_1 s_{2r+1}$ with $s_{2r+1} = f(s_0\ldots s_{2r})$ 
    and, by \textbf{(L4)}, we have $s_{2r+1} \to_2 s_{2r+2}$ with
    $s_{2r+2} = g(s_{2r+1})$ contradicting that $s_{2r}$ is the end
    configuration in the sequence $\sigma$. In conclusion, $g$ is indeed a 
    winning strategy for Player 2 (though it is \emph{not} a memoryless 
    strategy).
    By memoryless determinacy of 2-player reachability games, there exists a 
    memoryless winning strategy $g'$ for Player 2 reaching $F$ from $I_0$.
    \medskip

    \noindent
    ($\Rightarrow$) Let $A = post_{\to^*}(I_0)$.
    Consider a memoryless winning strategy $g: V_2 \to V_1$ of Player 2. 
    Such a strategy can be visualised as a forest $T$ whose set $V \subseteq
    S^*$ of nodes and set $E$ of edges
    are defined inductively as follows: (1) $A \subseteq V$, (2) if
    $w \cdot v_1 \in V$ for some $v_1 \in V_1\setminus F$, then 
    $w \cdot v_1v_2 \in V$ and $(w\cdot v_1,w\cdot v_1v_2) \in E$
    for every $v_2 \in V_2$ such that $v_1 \to_1 v_2$, and 
    (3) if $w \cdot v_2 \in V$ for some $v_2 \in V_2$, then 
    $w \cdot v_2v_1 \in V$ and $(w\cdot v_2,w\cdot v_2v_1) \in E$
    where $v_1 = g(v_2)$. 
    Since $g$ is winning, the height of each tree in the forest is finite
    and that each leaf in $T$ is in $S^*F$. In fact, each configuration $\pi$ in
    $T$ is a simple path (i.e. no node $s \in S$ appearing twice in $\pi$);
    for, if not, since $g$ is memoryless, we can construct a new strategy for 
    Player 1 by indefinitly 
    repeating the cycle on $s$ resulting in a loss for Player 2, which
    contradicts that $g$ is a winning strategy. We define a relation $R
    \subseteq V_1 \times V_1$ as follows: $(v_1,v_1')$ if there exists a
    configuration $\pi$ in $T$ in which $v_1$ appears \emph{strictly} before 
    $v_1'$. The
    relation $R$ is clearly transitive. We claim that $(v_1,v_1) \notin R$
    for every $v_1 \in V_1$. If $(v_1,v_1) \in R$ for some $v_1 \in V_1$,
    then there must exist $v_1' \in V_1$ such that $(v_1,v_1') \in R$
    and $(v_1',v_1) \in R$. The former is witnessed by a configuration $\pi$
    in $T$, while the latter is witnessed by a configuration $\pi'$ in $T$.
    Without loss of generality, we may assume that $\pi$ ends in $v_1'$, while
    $\pi'$ ends in $v_1$. Since $g$ is memoryless, as before we may construct 
    a new strategy for Player 1 as follows: from $\pi$, follow the suffix
    $v_1'\cdots v_1$ in $\pi'$, follow the suffix $v_1'\cdots v_1$, and 
    repeat this \emph{ad infinitum}. This contradicts the fact that $g$ is
    a winning strategy. In conclusion, $R$ is also irreflexive, which altogether
    implies that $R$ is a strict preorder. 

%% file: app-cex.tex
\section{Encoding of Counterexamples from
  Sect.~\ref{sec:counterexamples}}
\label{app:cex}

For $\mathit{CE}_1$, we introduce Boolean variables~$e_{i,q}$ for each
$i \in \{0,\ldots,|x|\}$ and state~$q \in \controls_A$, which will be
used to identify a path accepting $x$ in the automaton. We add
constraints that ensure that at least one $e_{i,q}$ is set for each
position~$i \in \{0,\ldots,|x|\}$, that the path starts at the initial
state~$q^A_0 = 1$ and ends in an accepting state, and that
the transitions on the path are enabled:
\begin{gather*}
  \Big\{ \bigvee_{q \in \controls_A} e_{i,q} \Big\}_{i \in \{0,\ldots,|x|\}},
  \quad
  e_{0, 1},\quad
  \big\{ \neg e_{|x|, q} \vee z^A_q \big \}_{q \in \controls_A},
  \\
  \Big\{
  \neg e_{i-1,q} \vee \neg e_{i, q'} \vee x_{(q, x_i, q')}
  \Big\}_{\substack{i \in \{1,\ldots,|x|\}\\ q, q' \in \controls_A}}
  ~.
\end{gather*}
In the last constraint, $x_i \in \ialphabet$ is the $i$th letter of $x$.

The encoding of $\mathit{CE}_2$--$\mathit{CE}_3$ is very similar to
the one of $\mathit{CE}_1$; at this point, it is important that the
automata are deterministic, since non-membership cannot be expressed
succinctly for NFAs.

\medskip
$\mathit{CE}_4$ is the most complicated counterexample, due to the
quantifier~$\exists z$. Since we assume that considered arenas are
length-preserving, $y \to_2 z$ implies that the length $|z|$ is known
from the counterexample. We can therefore introduce auxiliary
variables~$s_i$ ranging over $\ialphabet$ for each $i \in \{1, \ldots,
|z|\}$ to represent the letters of $z$ (the $s_i$ can be translated to
bit-vectors for the purpose of SAT solving). The individual atoms
$A(x)$, $A(z)$, $y \to_2 z$, $x \succ z$ can then be
translated separately to Boolean constraints, and combined to form
$\mathit{CE}_4$.

$A(x)$, $A(z)$, and $x \succ z$ can be encoded in a similar way as for
$\mathit{CE}_1$.  To encode~$y \to_2 z$, we represent the set $\{w \in
\ialphabet^* \mid y \to_2 w\}$ as an automaton
$(\ialphabet,\controls_S,\delta_S,q^0_S,F_S)$ (ideally a minimal one),
and again introduce Boolean variables~$e_{i, q}$ for each $i \in
\{0,\ldots,|z|\}$ and state~$q \in \controls_S$ to identify a path
accepting the word~$z$ in this automaton. The constraints resemble
those for $\mathit{CE}_1$:
\begin{gather*}
  \Big\{ \bigvee_{q \in \controls_A} e_{i,q} \Big\}_{i \in \{0,\ldots,|z|\}},
  \quad
  e_{0, q_S^0},\quad
  \{ \neg e_{|z|, q} \}_{q \in \controls_S \setminus F_S},
  \\
  \Big\{
  \neg e_{i-1,q} \vee \neg  e_{i, q'} \vee s_i \not= a
  \Big\}_{\substack{i \in \{1,\ldots,|z|\},\\
      (q, a, q') \in (\controls_S \times \ialphabet \times \controls_S)
      \setminus\delta_S}}
  ~.
\end{gather*}
The last constraint expresses that, whenever the accepting path visits
state~$q$ at position~$i-1$ and $q'$ at position~$i$, the letter~$s_i$
at position~$i$ has to be such that a transition~$(q, s_i, q') \in
\delta_S$ exists.

\anthony{I'm quite confused about the very last paragraph. What does it mean
by \emph{minimal} but not necessarily deterministic automaton? Also, I'm
rather unclear about the constraints. Can we talk about this over Skype?
Thanks!!}
\philipp{better?}
